\documentclass[review]{elsarticle}
\usepackage{lineno,hyperref,graphicx,color}
\modulolinenumbers[5]
\journal{Nuclear Instruments and Methods}
\newif\ifcomment
\newif\ifextrafigs
\newif\ifdraft
\newif\ifnotnimref
\notnimreftrue
\newcommand{\com}[1]{\relax}
\newcommand{\todo}[1]{\color{red}TODO: #1\color{black}}
\newcommand{\rem}[1]{{\footnote{\color{red}Remark: #1\color{black}}}}
\newcommand{\Fig}[1]{Fig.~\ref{#1}}
\newcommand{\Sec}[1]{Sec.~\ref{#1}}

\newcommand{\pt}{\ensuremath{p_\mathrm{T}}}

\begin{document}
\ifnotnimref
\begin{frontmatter}
\title{Design and performance of a silicon tungsten calorimeter prototype module and the associated readout}
\address[ornl]{Oak Ridge National Laboratory, Oak Ridge, Tennessee, United States}
\address[tsukuba]{University of Tsukuba, Tsukuba, Japan}
\address[tsukuba_tech]{Tsukuba University of Technology, Tsukuba, Japan}
\address[bergen]{Department of Physics and Technology, University of Bergen, Bergen, Norway}
\address[cns]{Center for Nuclear Study, University of Tokyo, Tokyo, Japan}
\address[utrecht]{Institute for Subatomic Physics of Utrecht University, Utrecht, Netherlands}
\address[lund]{Lund University, Sweden}
\address[nara]{Nara Women's University, Nara, Japan}
\address[hiroshima]{Hiroshima University, Higashi Hiroshima, Japan}
\author[ornl]{T.~Awes}
\author[ornl]{C.L.~Britton}
\author[tsukuba]{T.~Chujo\corref{mycorrespondingauthor}}
\cortext[mycorrespondingauthor]{Corresponding author}
\ead{chujo.tatsuya.fw@u.tsukuba.ac.jp}
\author[ornl]{T.~Cormier}
\author[ornl]{M.N.~Ericson}
\author[ornl]{N.B.~Ezell}
\author[bergen]{D.~Fehlker}
\author[ornl]{S.S.~Frank}
\author[tsukuba]{Y.~Fukuda}
\author[cns]{T.~Gunji}
\author[nara]{T.~Hachiya}
\author[cns]{H.~Hamagaki}
\author[cns]{S.~Hayashi}
\author[tsukuba]{M.~Hirano}
\author[tsukuba]{R.~Hosokawa}
\author[tsukuba_tech]{M.~Inaba}
\author[tsukuba]{K.~Ito}
\author[tsukuba]{Y.~Kawamura}
\author[tsukuba]{D.~Kawana}
\author[tsukuba]{B.~Kim}
\author[tsukuba]{S.~Kudo}
\author[ornl]{C.~Loizides}
\author[tsukuba]{Y.~Miake}
\author[utrecht]{G.~Nooren}
\author[tsukuba]{N.~Novitzky}
\author[utrecht]{T.~Peitzmann}
\author[ornl]{K.F.~Read}
\author[bergen]{D.~R\"{o}hrich}
\author[nara]{T.~Sakamoto}
\author[tsukuba]{W.~Sato}
\author[cns]{Y.~Sekiguchi}
\author[nara]{M.~Shimomura}
\author[ornl,lund]{D.~Silvermyr}
\author[ornl]{P.W.~Stankus}
\author[hiroshima]{T.~Sugitate}
\author[tsukuba]{T.~Suzuki}
\author[hiroshima]{S.~Takasu}
\author[utrecht]{A.~van den Brink}
\author[utrecht]{M.~van Leeuwen}
\author[bergen]{K.~Ullaland}
\author[utrecht]{H.~Wang}
\author[ornl]{R.~Warmack}
\author[bergen]{S.~Yang}
\author[utrecht]{C.~Zhang}
\date{\today}
\begin{abstract}
We describe the details of a silicon-tungsten prototype electromagnetic calorimeter module and associated readout electronics. 
Detector performance for this prototype has been measured in test beam experiments at the CERN PS and SPS accelerator facilities in 2015/16.
The results are compared to those in Monte Carlo Geant4 simulations.
This is the first real-world demonstration of the performance of a custom ASIC designed for fast, lower-power, high-granularity applications.
\end{abstract}
\end{frontmatter}
\ifdraft
\linenumbers
\fi
\section{Introduction}
\label{sec:intro}
The parton structure of protons and nuclei is typically characterised in terms of parton distribution functions~(PDFs) which parameterise the non-perturbative physics that cannot at present be calculated from Quantum Chromodynamics (QCD)~\cite{Wilczek:2000ih,qcdbook}.
The PDFs are determined from global fits to experimental measurements, in particular from deep inelastic scattering~(DIS) experiments such as H1 and ZEUS at HERA~\cite{Abramowicz:2015mha}.
The gluon PDF is found to rise dramatically in the small-$x$ region~($x = 10^{-2} - 10^{-6}$), 
where the Bjorken variable $x$ is the longitudinal fraction of the momentum of the nucleon carried by the partons (quarks and gluons).
At small $x$ one expects non-linear behaviour of QCD with competing processes of gluon splitting and fusion, which should eventually lead to saturation of the gluon density.
Despite extensive experimental studies, there is no direct evidence of gluon saturation, nor the creation of the Color Glass Condensate~\cite{Iancu:2003xm}.
By measuring direct photons in the forward direction, which are produced in the quark-gluon Compton process, one directly probes the gluon PDF, which allows exploring QCD in the non-linear regime, and placing stringent constraints on the gluon nuclear PDFs~\cite{vanLeeuwen:2019zpz}. 

Motivated by the considerations above, a proposal was developed to build a forward calorimeter system in the ALICE experiment at the LHC, called FoCal~\cite{Peitzmann:2018kie,LoI}.
The FoCal is composed of an electromagnetic part (FoCal-E) and hadronic part (FoCal-H), with a planned pseudo-rapidity coverage of about $3.4 \le \eta \le 5.8$. 
The FoCal-E is designed to have an excellent two-shower separation to identify and reject 
decay photons from neutral mesons with transverse momenta in the range from a few GeV/$c$ to above 20 GeV/$c$, i.e.\ total momenta up to about 1 TeV/$c$ ($\pt = 5$ GeV/$c$ corresponds to $p\approx370$ GeV/$c$ at $\eta = 5$) to  enable the measurement of direct photons at the LHC in the small-$x$ region.
It uses tungsten converter layers with a thickness of 1 radiation length each and silicon sensor layers, with a compact design to minimise the shower size. 
The sensor layers consists of silicon pad sensors with a granularity of $1\times1$~cm$^2$ (low granularity) that will primarily measure the energy of the shower, and a few high-granularity pixel layers that will be used in the future detector to split contributions from overlapping showers. \footnote{It should be noted that both types of sensor layers are of fine granularity in comparison to conventional calorimeters. We use the terms \textit{low} and \textit{high} granularity in this context only to discriminate further between the two types of layers in our calorimeter prototype.}

A Si-W calorimeter prototype using the Si-Pad technology of similar granularity~($1\times1$~cm$^2$ and more recently $0.5\times0.5$~cm$^2$) has previously been built and tested by the CALICE collaboration~\cite{collaboration_2008,Adloff_2009,Kawagoe:2019dzh}. 
The prototype discussed here uses a different approach for readout electronics, summing signals from several layers to reduce the necessary number of channels without compromising the performance needed for measurements of electromagnetic showers. 
In addition, the use of pixel layers with their much higher granularity is a unique feature. The analysis of events with simultaneous read out of the high-granularity and low-granularity layers shows that sub-millimeter precision on the shower position is obtained with this setup. 

\begin{figure}[t!]
\begin{center}
 \includegraphics[width=100mm,clip]{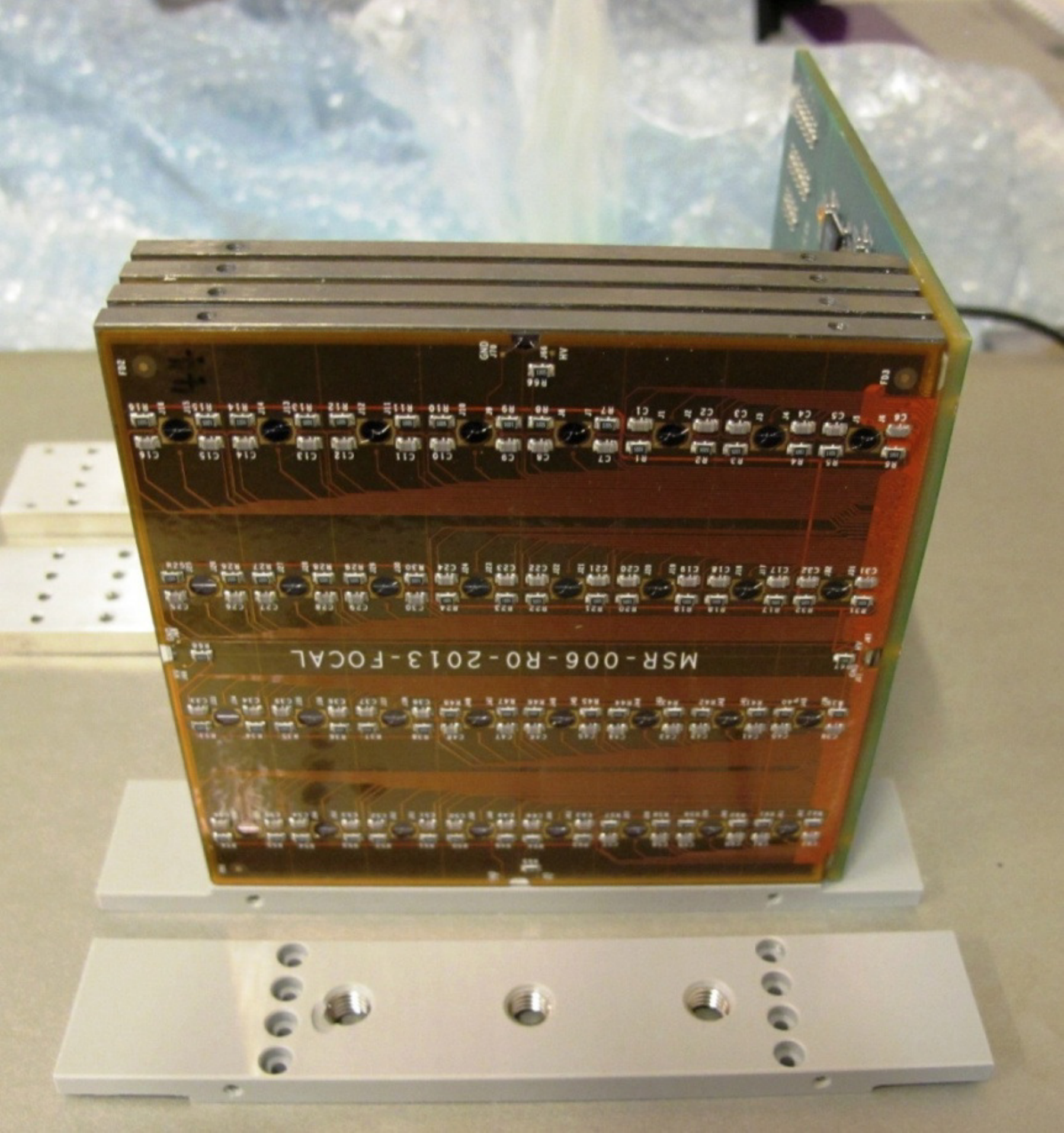}
\end{center}
\vspace{-5mm}
\caption{A FoCal PAD detector segment. The silicon layer and PCB are in front. The analog summing circuit board is shown on the right.}
\label{fig:pad_pic}
\end{figure}

In this paper, we present the results of test beam measurements at the CERN Proton Synchrotron~(PS) and the Super Proton Synchrotron~(SPS) in 2015/2016 for a prototype using pad and pixel layers. 
The pad readout uses a new custom ASIC developed for very fast~(fast enough for low-level trigger decision), lower power, fine granularity~(reduced cost per channel) applications such as the FoCal. The ASIC provides differential drive to downstream electronics which can be located farther from the acceptance with reduce requirements for radiation tolerance.
For the test setup with only pad layers, the energy resolution, energy linearity, and shower profiles are presented and compared to simulations using the Geant4 simulation took kit~\cite{AGOSTINELLI2003250}, version 10.5.01 with physics lists FTFP and BERT.
We also show the performance of the integrated system, i.e.\ combined pad and pixel layers together, as a first-principle design of the FoCal-E prototype from the 2016 test beam data.

\section{Detector design}
\label{sec:design}
The FoCal-E design is designed to provide very high lateral segmentation to discriminate between decay photons and direct photons.
The electromagnetic calorimeter requires compact shower size to minimize the effect of shower leakage and optimize shower separation.
Therefore, the FoCal-E prototype has been designed as a silicon and tungsten (Si+W) sampling calorimeter, because tungsten, which is used as an absorber material, has a small Moli\`ere radius of 9~mm and a radiation length of 3.5~mm. 
The prototype of FoCal-E has two types of Si sensors: pad and pixel layers.

\begin{figure}[t!]
\begin{center}
\includegraphics[width=100mm,clip]{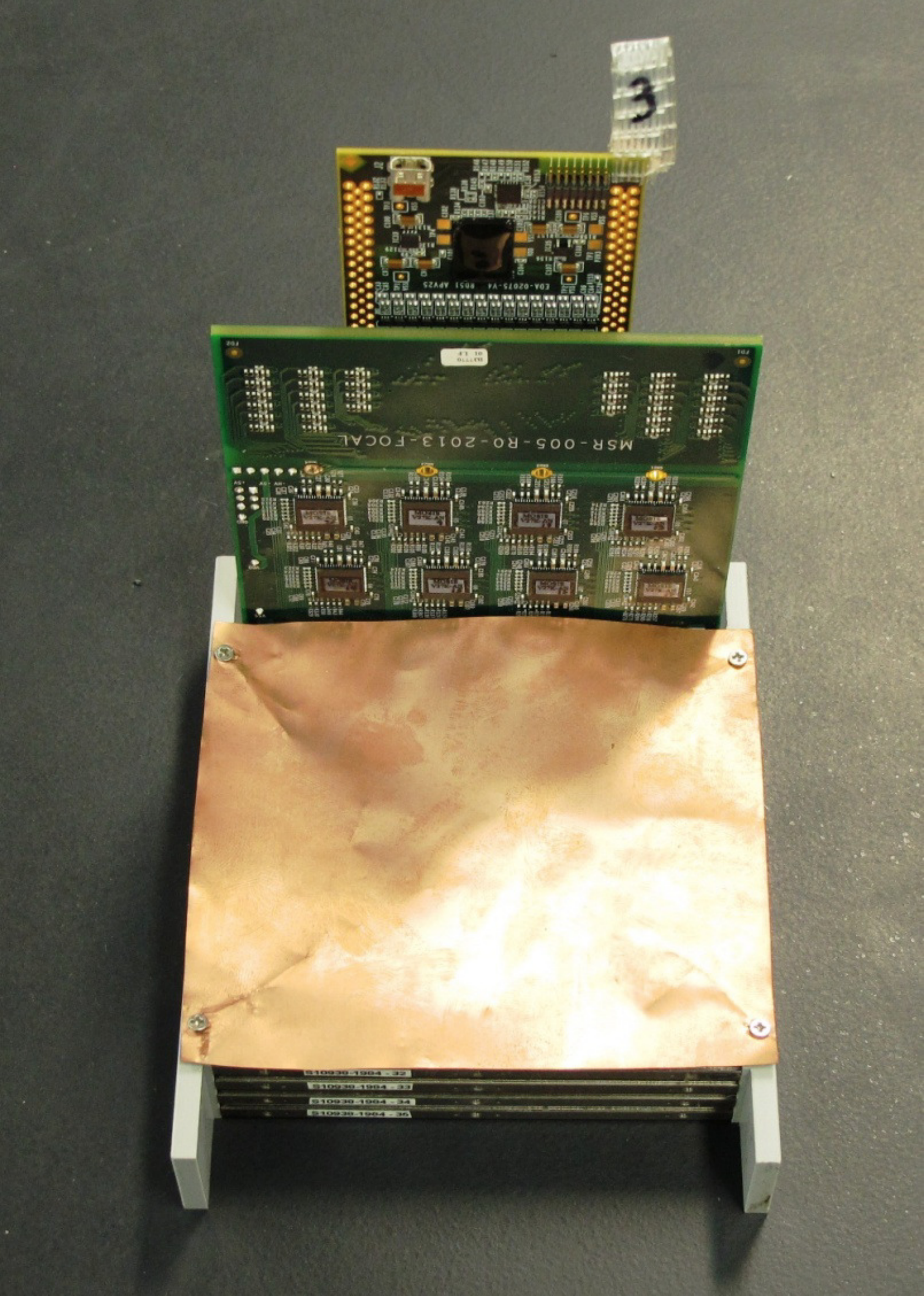}
\end{center}
\vspace{-5mm}
\caption{A FoCal PAD detector segment with the protection cover and APV25 hybrid front-end card  attached on the summing board.}
\label{fig:pad_cupper}
\end{figure}

For the pad layer, we used a 64 pad silicon sensor, with 93~mm$^2$  overall pads, type Hamamatsu S10938-9959, arranged as 8 $\times$ 8 cells with 11.30~mm pitch.
The front-end electronics use a new custom ASIC developed for this (and other) application(s) where fast, fine-granularity imaging is required. 
This FoCal prototype is the first real-world application for this new ASIC. 
The ASIC is located on the summing board. 
The output from the summing board was sent to a test beam data acquisition system. 
In a final, full-scale application alternative ASICs receive the output of the summing board~(such as the Beetle chip or an APV25 hybrid as described in \Sec{sec:readout}).

The assembled detector consists of four segments, each of which contain 4 layers consisting of Si pad sensor mounted using transfer tape (0.2~mm) onto a W plate~(3.5~mm thickness, 1 $X_0$), as shown in \Fig{fig:pad_pic}.
An analog signal from each pad layer is read out through 0.05~mm aluminum wires bonded to the flexible printed circuit board (PCB), also mounted with transfer tape to the sensor, and the analog signals from the PCB with the same pad position on 4 layers are summed longitudinally on the analog summing board. 
The sum forms the signal of a \textit{cell}. 
For the readout system after the summing of signals, we use an APV 25 hybrid board as a front-end card and the SRS (Scalable Readout System ) developed by the CERN RD51 collaboration. 
The mmDAQ system is used for the DAQ control of the SRS system. 

\Fig{fig:pad_cupper} shows one FoCal pad detector segment with the Si layer, PCB, analog summing circuit board, and APV25 hybrid front end card.
This detector configuration (with a total of four of this kind of segment) is used for the CERN test beam experiments.

\Fig{fig:pad_ps_sps} shows the detector setup of the test beam experiment at PS and SPS at CERN in 2015. 
Four pad detector segments were aligned in an aluminum rail to fix the detector position. 
The beam enters from right to left~(starting from the segment labelled ``FoCal 1'' to``FoCal 4'' in \Fig{fig:pad_ps_sps}).
Four independent APV25 hybrid boards were attached on the summing board and read out by the SRS system, which consists of a Front-end-Card (FEC) and an Analog-to-Digital Converter~(ADC).
In the following, we label the pad sensor segments FoCal~1 to FoCal~4 as low granularity layer (LGL) with segment numbers starting from 0 (LGL0 to LGL3).

\begin{figure}[t!]
\begin{center}
\includegraphics[width=130mm,clip]{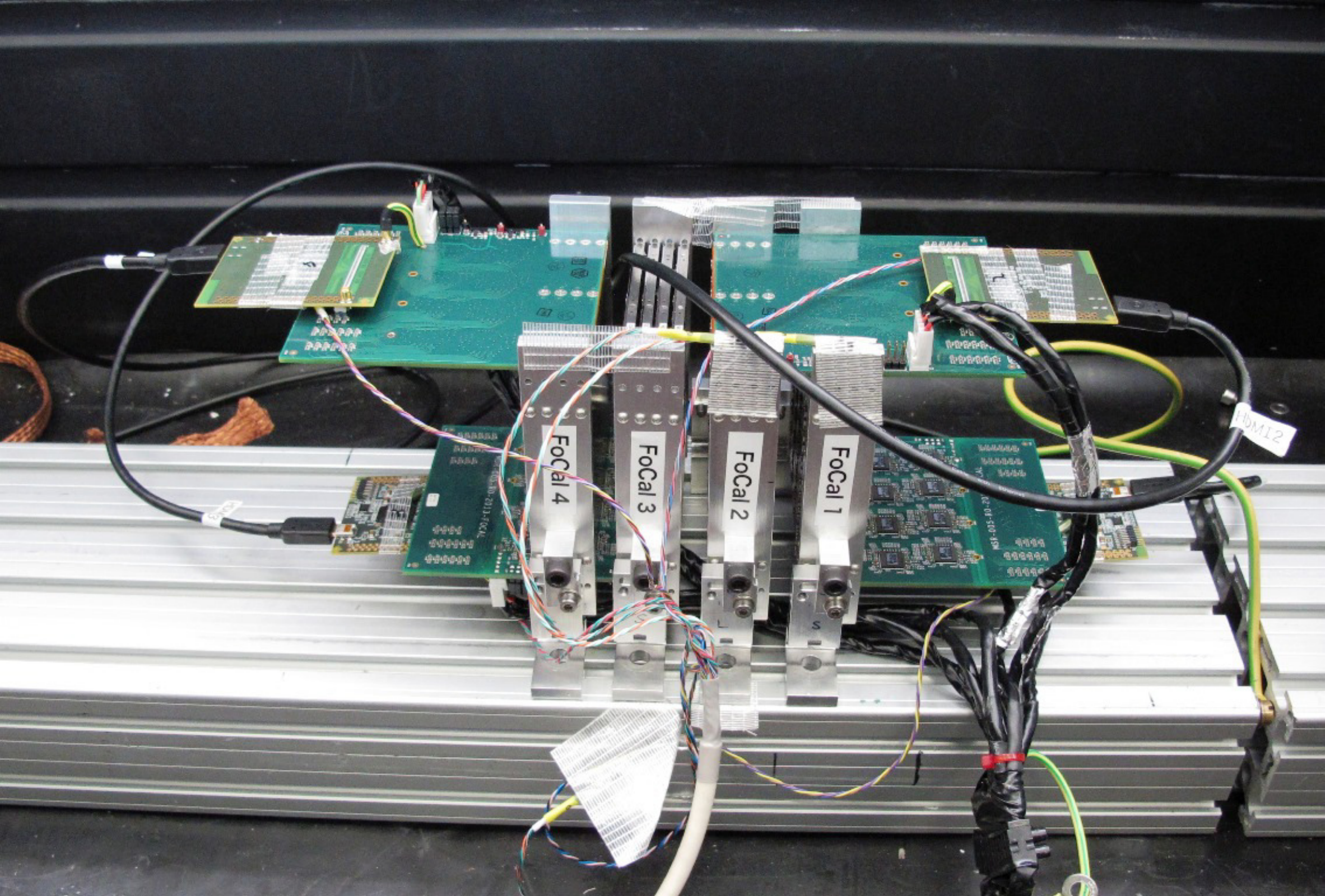}  
\end{center}
\vspace{-5mm}
\caption{Four FoCal PAD detector segments for the PS/SPS test beam.}
\label{fig:pad_ps_sps}
\end{figure}

Two layers of pixel sensors were inserted between the first and second pad segment (LGL0 and LGL1) and between the second and third segment (LGL1 and LGL2). The design of these layers is based on a prototype of a fully digital pixel calorimeter as described in \cite{de_Haas_2018}. 

The pixel layers use MIMOSA23 sensor chips from IPHC \cite{phase1}, which have a size of $19.52 \times 20.93\; \mathrm{mm^2}$.
For each chip, $19.52 \times 19.52\;\mathrm{mm^2}$ is occupied by the pixel matrix, consisting of $640 \times 640$ pixels.
In this prototype, a pixel layer consists of four individual chips. 
The resulting lateral size of a pixel layer for this test is $38.4 \times 38.4\;\mathrm{mm^2}$ with $1280 \times1280$ pixels. 
Each sensor has 640 discriminators, one per pixel column, which produces a binary readout. The threshold of the discriminators can be set externally and has been optimised to reach good sensitivity with low noise ($< 10^{-5}$ noise rate per pixel). 
The pixel sensors are controlled by a dedicated readout system \cite{Fehlker:2013dga}, which consists of two levels of FPGAs. 
The sensor itself is read out in a 'rolling shutter' mode, in which the pixel data are digitised and shipped row by row with a frequency of 1 MHz, resulting in a total readout time of 640~$\mu$s for the full sensor. 
We utilise a so-called 'full-frame readout', i.e.\ all digitised pixel values are read out independent of their value and no zero suppression is performed. 
The main tasks of the FPGAs are to multiplex and buffer the data before shipping them to a PC over Gigabit Ethernet. 
While the readout rate of the sensor is adequate for a test beam, a faster sensor will be needed for the final application in the FoCal. The ALPIDE chip~\cite{Mager:2016yvj} of the new ALICE inner tracking system is a clear candidate, which was still being developed at the time the data presented here were collected.

Because the data acquisition systems for the pixels and pads were independent, the data streams were synchronized between the two systems with a common event identifier generated by a trigger counter. 

\section{Readout electronics}
\label{sec:readout}
Traditional charge-sensitive preamplifiers~(CSP) are commonly used for readout of ``capacitive'' detectors~(detectors such as silicon pads, strips, etc.\ for a which a simplified model consists of a capacitor in parallel with a current generator) for two reasons. 
First, all the charge generated in a detector due to a radiation event is ultimately collected by the preamplifier irrespective of the detector capacitance.
Higher detector capacitance tends to slow the preamplifier bandwidth such that it may take many microseconds to collect the charge but it will ultimately be collected. 
Second, the ratio of the output voltage to the input charge (charge gain) is determined by the feedback capacitor used in the CSP and not by the detector. 
Since $Q/C=V$, this will allow a small charge signal to be processed by a small feedback capacitor on the CSP instead of that same small charge on a much larger detector capacitance.
This results in a proportionally larger voltage signal for subsequent processing.

Because of the large amount of charge/event available in the FoCal PAD detector and the need for a fast trigger signal (fast preamplifier response), a traditional CSP is likely not ideal or needed. 
Therefore, a truly application-specific approach to on-chip readout was developed using a new custom ASIC developed with FoCal as the leading use-case. 
The ASIC provides a fast, low power and low cost per channel solution that satisfies the requirements for our application (and others requiring cost-effective tiling of large areas with high granularity) without the extra complexity, cost, or power of maintaining a closed-loop CSP.
The noise for this follower topology will be competitive with that for a well-designed (and more complicated) CSP, but at a typically lower cost per channel and lower power per channel.  
As a dimensional argument for a generic detector sensor, we note the detector capacitance is inversely proportional to the thickness as given by
\begin{equation}
C=\frac{A}{T}\cdot\epsilon_0\epsilon_r
\end{equation}
where $C$ is the plate-plate capacitance, $A$ is the area of the detector plates, $T$ is the thickness of material, and $\epsilon_0$ and $\epsilon_r$ are the absolute and relative dielectric constants of the material, in this case silicon. 
The amount of charge per event is
\begin{equation}
Q \propto A\cdot T\cdot P(T)
\end{equation}
where $P$ is the linear amount of energy deposited in a thickness $T$ of material.
Since $Q/C = V$, we obtain for the voltage out of the detector for a given pulse
\begin{equation}
V_{\rm out}\propto \frac{T^2\cdot P(T)}{\epsilon_0\epsilon_r}
\end{equation}

As a consequence of these general considerations, we can optimize the detector electronics for the desired noise and gain by specifying the detector thickness, a very useful attribute indeed. 
This was done to a limited extent in this first version of the custom ASIC reported here~(with further optimization possible in subsequent revisions).
For simplicity, we utilized a very fast high-speed follower topology similar to that used on a photomultiplier tube.
This allows us to maintain high speed, low noise and simplicity at the front end detector. 
With a follower, we have sufficient bandwidth to provide a fast trigger without having to maintain a high bandwidth, closed-loop CSP.
Processing electronics can be placed away from the detector, thereby somewhat mitigating heat and power-distribution problems.
The follower is the ultimate in simplicity.
Simulations show that if we are able to design the detector and follower circuit such that our input maximum charge results in approximately 1.6~V output, we can develop a circuit which will exhibit noise of approximately 108~$\mu$V rms, a peak/rms ratio of 14,800.
This shows that we will likely not be limited by noise but by inter-pad cross talk.

\begin{figure}[t!]
\begin{center}
\includegraphics[width=10cm]{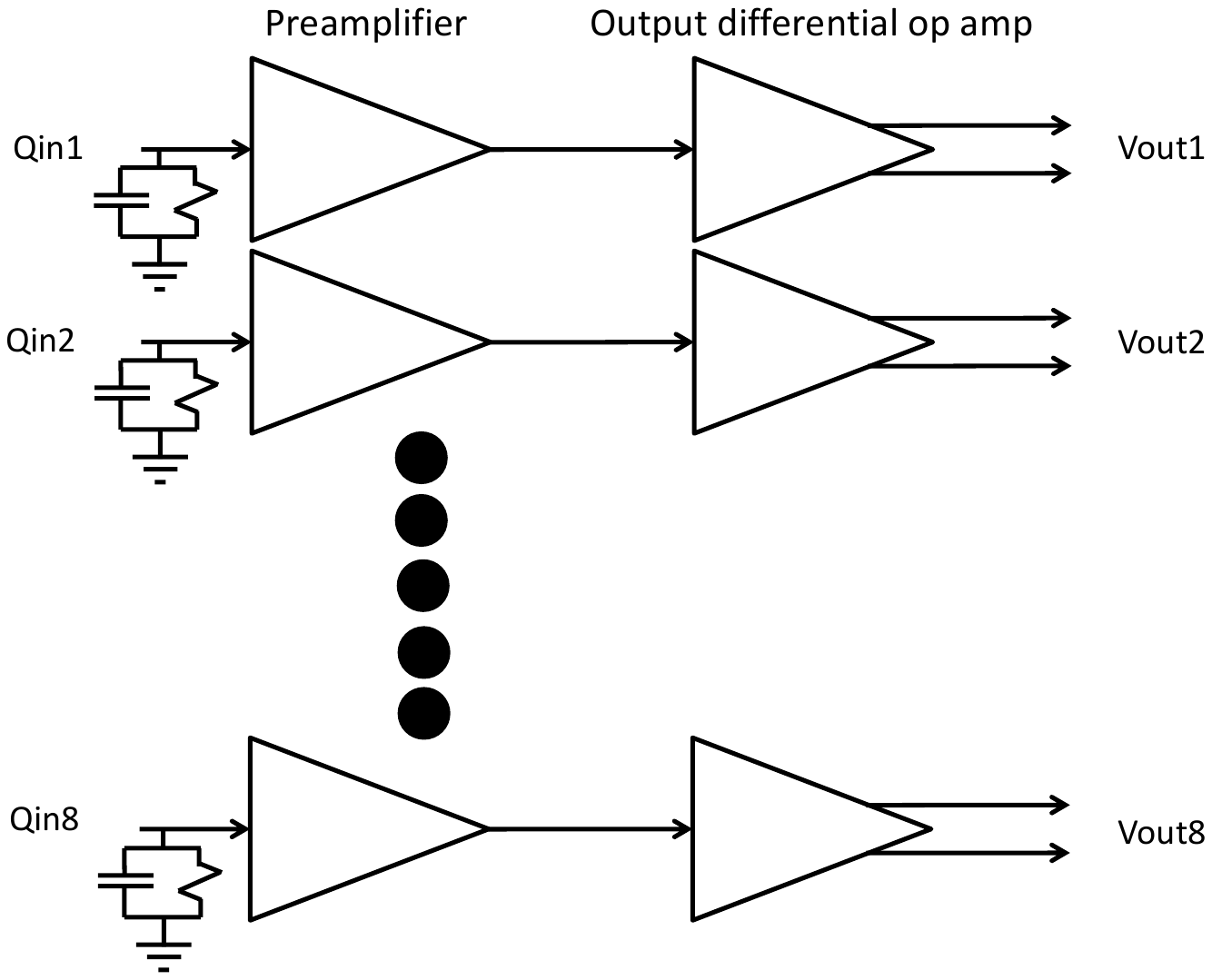}
\end{center}
\vspace{-5mm}
\caption{Preamplifier buffer ASIC block diagram.}
\label{fig:preampblock}
\end{figure}

The follower circuit requires a buffered output, preferably differential, to minimize cross talk (as shown in \Fig{fig:preampblock}). 
The output of the differential buffer drives the signal to an area with more available space, where it can be connected to processing electronics (shaper, trigger processor, ADC), simplifying their requirements.

\begin{figure}[t!]
\begin{center}
\hspace*{-1cm}\includegraphics[width=14cm]{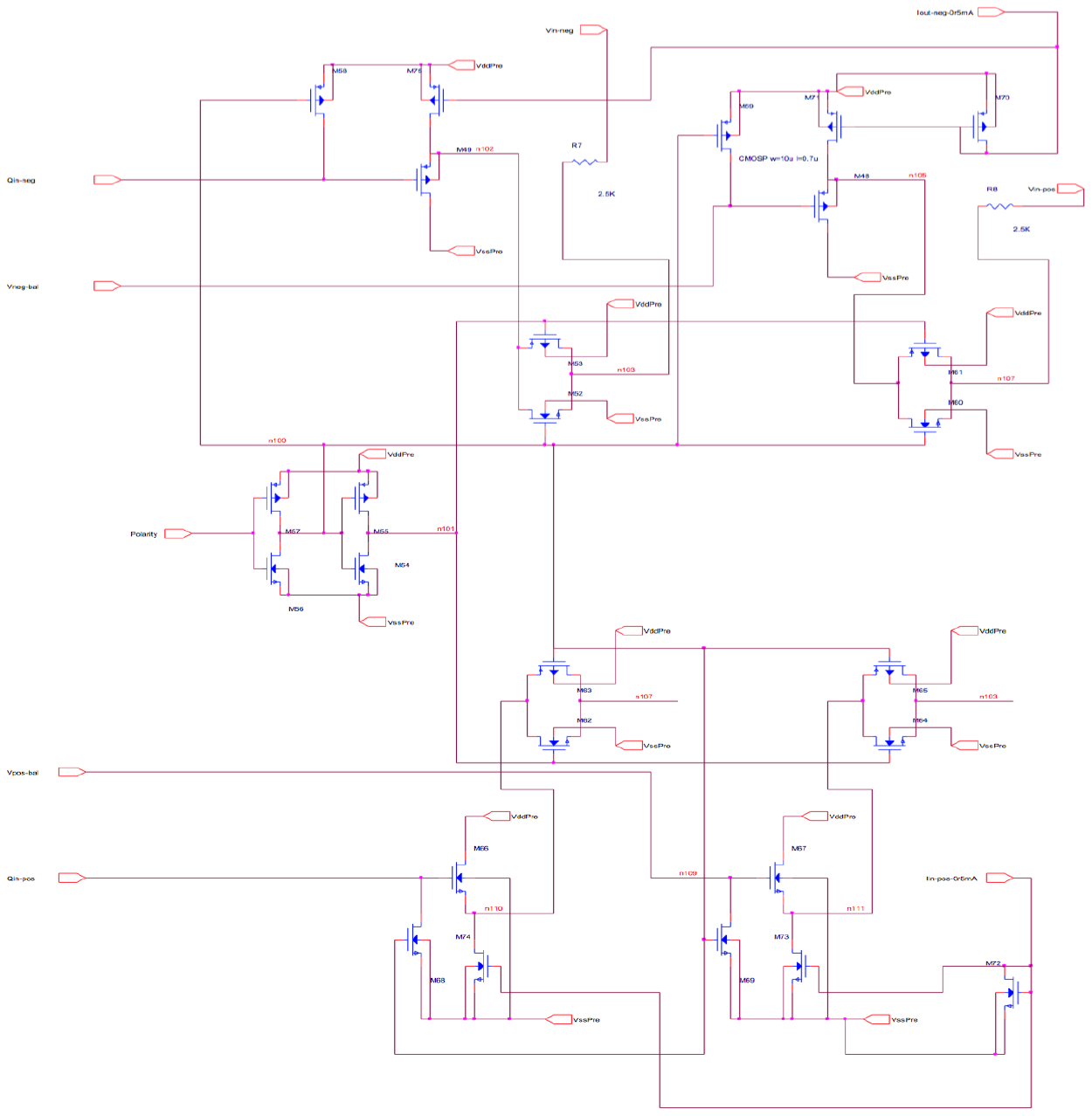}
\end{center}
\vspace{-2cm}
\caption{Preamplifier schematic drawing showing negative follower and dummy level adjust (upper portion) and positive follower and dummy level adjust~(lower portion).}
\label{fig:preampschem}
\end{figure}

\begin{figure}[t!]
\begin{center}
\hspace*{-2cm}\includegraphics[width=12cm]{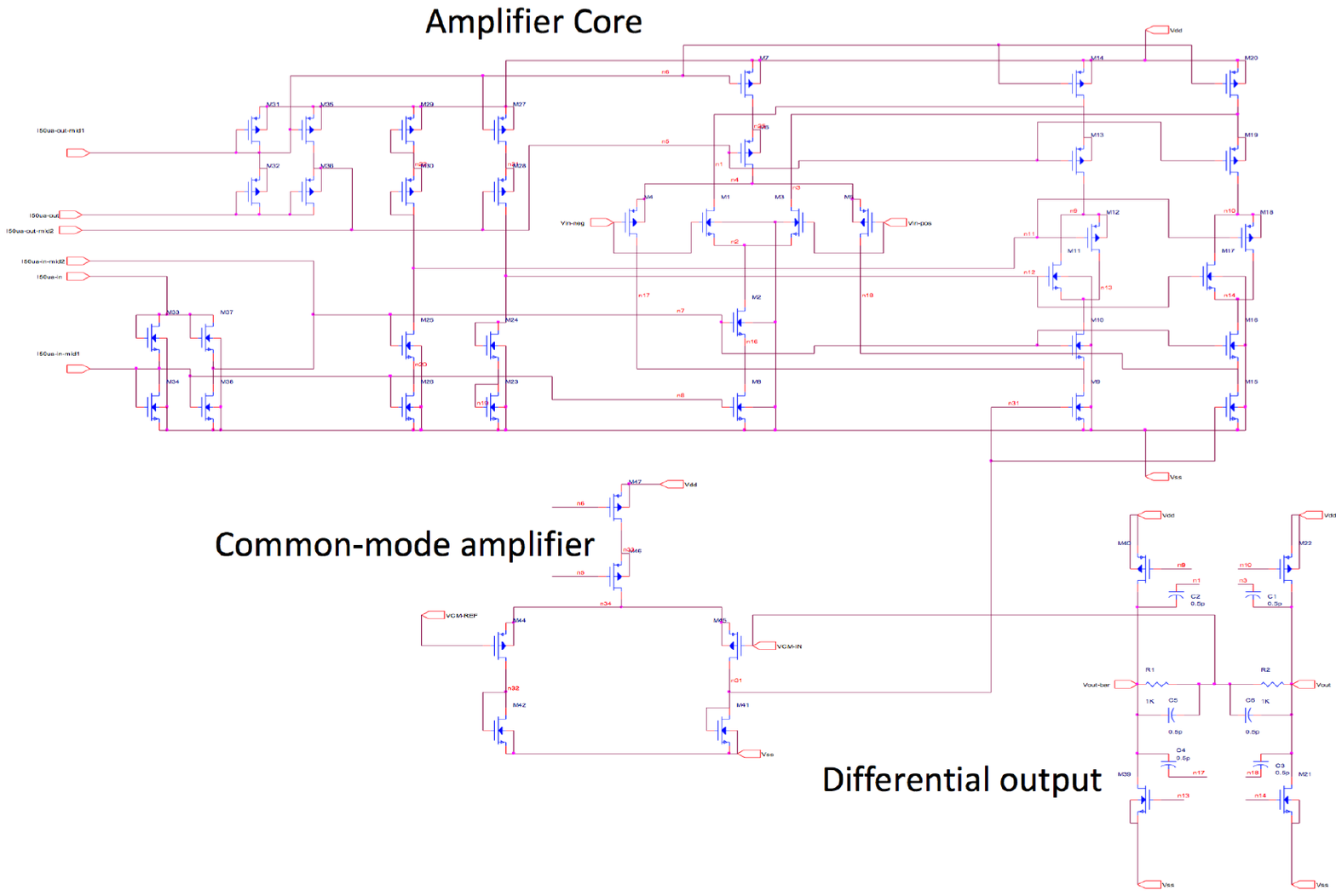}
\end{center}
\vspace{-5mm}
\caption{Output differential operational amplifier showing amplifier core, common-mode amplifier, and differential output.}
\label{fig:opamp}
\end{figure}

The preamplifier connects to the detector through a coupling capacitor, if needed, and can utilize either polarity of input charge.
There are bias setting resistors on the chip that set the quiescent input voltage.  
When an event occurs, the charge is collected on the detector capacitance and the voltage output is buffered and sent to the single-ended-to-differential driver.  
This driver is designed to drive a 100~ohm differential line.  
The power dissipation is currently under 10~mW for the entire circuit which operates on 2.5~V.  
The preamplifier was implemented using the On Semi 0.5 $\mu$m CMOS process.
The layout estimate results in a chip area of under $2\times2$~mm$^2$ for four channels.

The preamplifier-buffer ASIC block diagram is shown in \Fig{fig:preampblock}.  
The source follower preamplifier feeds the output differential amplifier.

The detailed preamplifier schematic is presented in \Fig{fig:preampschem}.  
It includes the source-follower inputs (one for each input polarity) and the switching and routing circuitry for each.  
For wide applicability across multiple potential detector types, we allow selection of either polarity. For each polarity, there are dummy devices used to maintain bias for the inputs to the differential drive amplifier.

\Fig{fig:opamp} shows the fully differential input/output operational amplifier schematic.  
This design implements a floating class A-B control buffer~(Hogervorst) as well as a common-mode control loop for the output.

\begin{figure}[thb!]
\begin{center}
\includegraphics[width=135mm,clip]{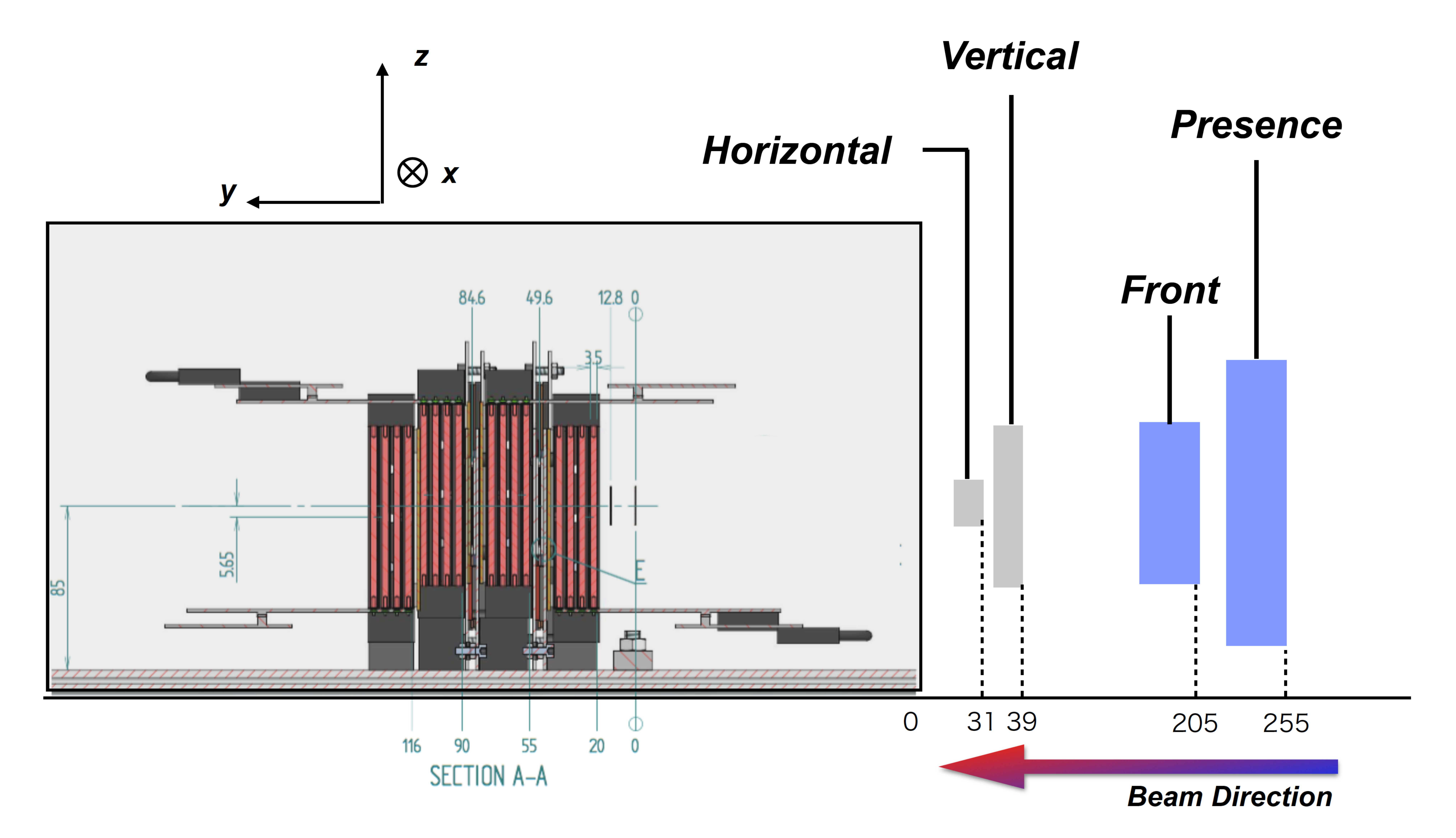}
\end{center}
\vspace{-5mm}
\caption{Test beam setup at PS and SPS.}
\label{fig:setup}
\end{figure}

\section{Detector setup for test beam experiments}
\label{sec:testbeam}
The test beam performance for the prototype detector was carried out at PS and SPS facilities at CERN in 2015 and 2016. 
In 2015, we used the T9 test beam facility at the PS accelerator providing a secondary beam (with mixed electrons and hadrons) with a beam energy from 0.5 to 10~GeV. 
In the same year we used the SPS T4-H6 beam line with the energy setup of 5--180~GeV. In 2016, the SPS H6B beam line was used with a very similar setup as the previous year.

The purity of the electrons or positrons in the PS T9 facility is around 80\% at lowest momenta and then rapidly dropping. 
Therefore, at beam energies higher than 1 GeV we used a Cherenkov trigger in order to increase the purity of the electrons in the beam. 
At SPS energies, the fraction of electrons in the beam was high ( $> 50 \%$) for energies up to 50 GeV, and decreased for larger energies.

\Fig{fig:setup} shows the prototype detector configuration in both years and test beam facilities. 
We combine the four segments of pad layers~(see \Sec{sec:pads}) with the two pixel layers~(see \Sec{sec:maps}) and put the complete system into a light-tight box for data taking. 
The first (second) layer of pixel sensors is located at 29.6 (64.6) mm from the front face of the full detector setup. 
These locations correspond to depths of 4 and 8 radiation lengths, respectively.
For the beam trigger, we used in total 4 different plastic scintillators: (1) of 10 $\times$ 10 ${\rm cm}^2$ area to detect the beam presence, (2) of 4 $\times$ 4 ${\rm cm}^2$ area with 1~cm thickness for the narrow coincidence, (3) and (4) of smaller size, which are placed horizontally and vertically to define together a 1$\times$ 1 ${\rm cm}^2$ fiducial trigger in the center of the active detector area.

In summary, for the data taking periods in 2015 we took data with beam energies from 0.5 to 5 GeV at PS and 10 to 50 GeV at SPS, while in 2016 we took data with beam energies of 50 to 130 GeV. 
For the latter period, a large dead area was observed in the pad readout, concerning, in particular, the LGL3 segment. Therefore, in this paper for these higher energy data the pad layers were not used, and only the pixel results are shown. 
For a sample of limited statistics, we were able to send the same trigger bit to both data streams, therefore we could combine it offline and match the pad- and pixel-layers. 
During these measurements the readout electronics of the pad-layers had to be configured for the high-gain setting, implying that for these tests the pads could measure only the hit position but not the energy.

\begin{figure}[tb!]
\begin{center}
\includegraphics[width=12cm]{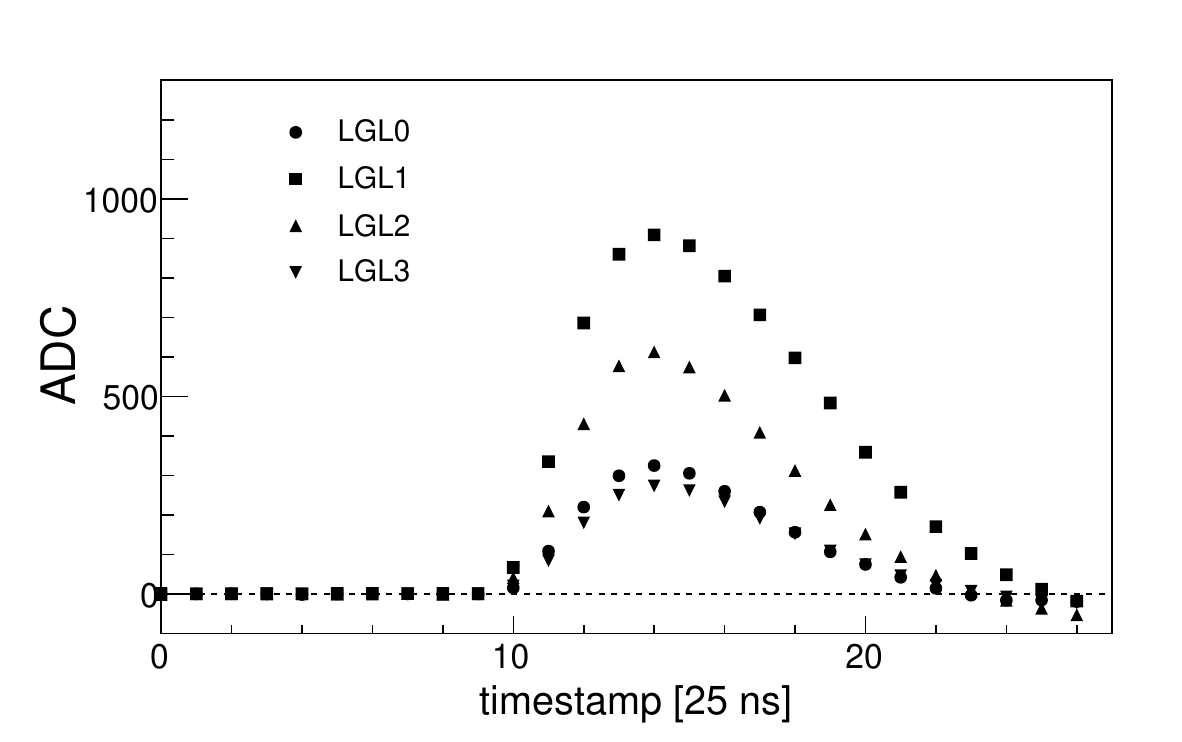}
\end{center}
\vspace{-5mm}
\caption{Measured charge (ADC) versus time for cells of different PAD segments (LGL0--LGL3) using 40 GeV/$c$ electron beams at SPS.}
\label{fig:amp_40gev}
\end{figure}

\section{PAD detector performance}
\label{sec:pads}
\label{sec:pad_performance}
\Fig{fig:amp_40gev} shows the distribution of measured charge (ADC) for single cells (i.e. longitudinally summed groups of four consecutive pads) as a function of time for the four segments (LGL0, LGL1, LGL2 and LGL3) for~40 GeV positrons.
The horizontal axis is time~(25~ns per bin), and the vertical axis is the charge collected at the summing board on each LGL, i.e.\ average amplitude of the ADC values over all events in each time bin.
Of the four segments, LGL1 shows the largest signal, which corresponds to the location of the longitudinal maximum of the shower, as expected for 40 GeV beam energy.

The peak value of these cell signals is then digitized. To suppress contributions from background, we performed pedestal and common mode noise subtractions and remove signals that are below 4 times the width of the pedestal peak.
Furthermore, we assure that the beam hit positions on all four pad layers have the same $x$ and $z$ positions (perpendicular to the beam direction), by looking at the electromagnetic shower signal on each pad layer. 
We apply this \textit{directional cut} in order to select only beam particles travelling parallel to the $y$-direction. 
Offline, we performed 3$\times$3 cell clustering in each segment to reconstruct EM showers, and the cluster signals of all segments are added.

To summarise, the total ADC sum has been calculated by the following steps:
\begin{enumerate}
    \item The analog signals of pads at the same transverse location within a segment are summed longitudinally (in hardware using the summing boards) to obtain cell signals.
    \item The cell signals are sampled with the ADC around the maximum time bin, and the digitised value is stored.
    \item Shower clustering using 3 $\times$ 3 cells is performed offline for each LGL segment.
    \item A sum over the four LGL segments is performed.
\end{enumerate}

\begin{figure}[t!]
\begin{center}
\includegraphics[width=12cm]{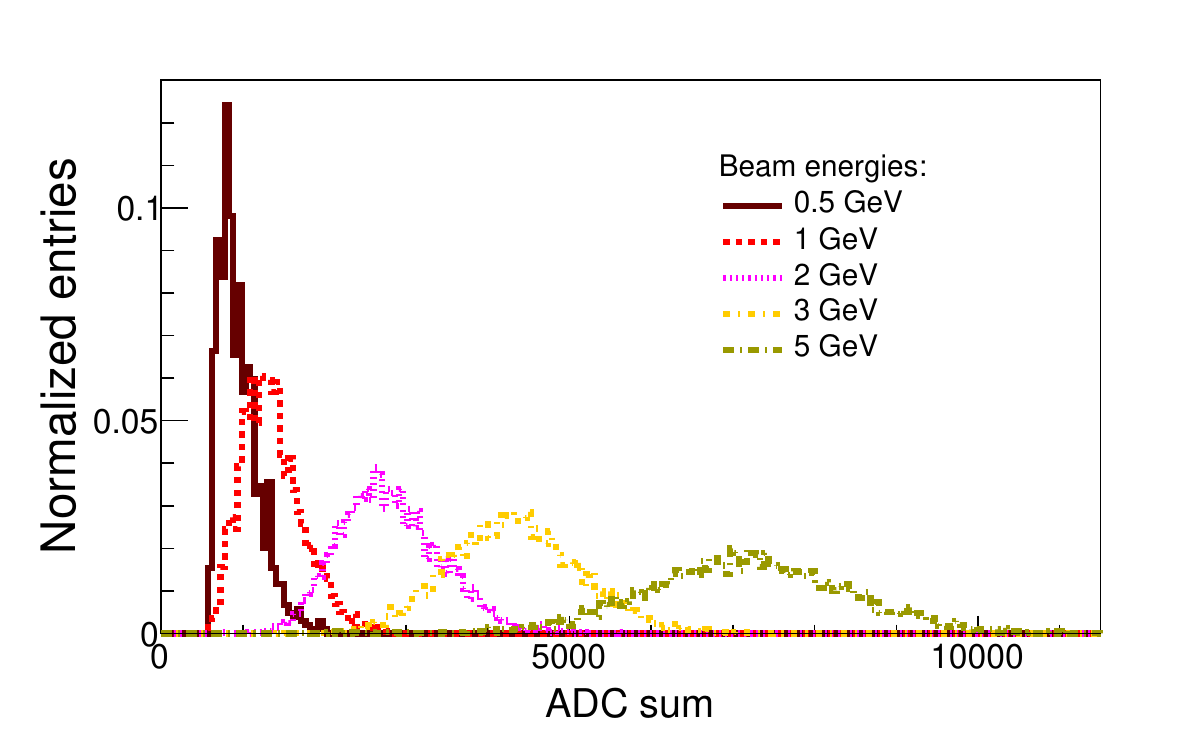}
\end{center}
\vspace{-5mm}
\caption{ADC sum distribution from 0.5 to 5 GeV electron beams at the PS.}
\label{fig:adcsum_PS}
\end{figure}
\begin{figure}[b!]
\begin{center}
\includegraphics[width=12cm]{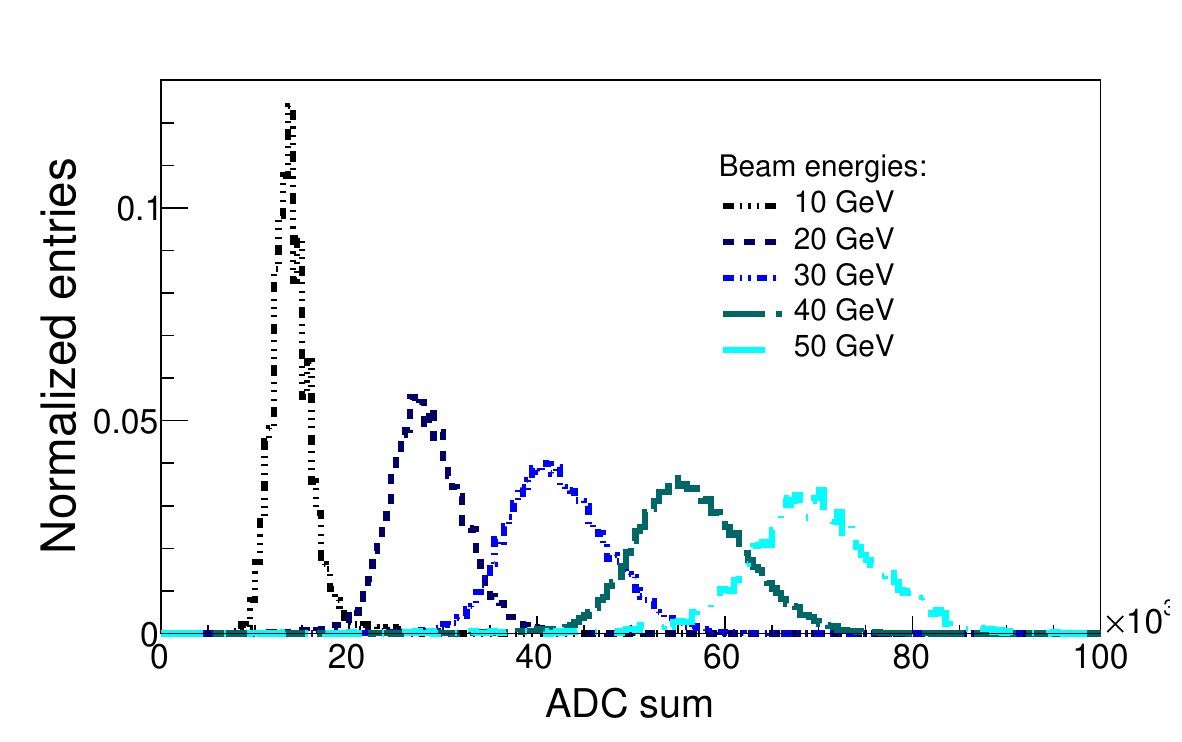}
\end{center}
\vspace{-5mm}
\caption{ADC sum distribution from 10 to 50 GeV electron beams at the SPS.}
\label{fig:adcsum_SPS}
\end{figure}

\Fig{fig:adcsum_PS} shows the distribution of the total ADC sum for electrons from 0.5 to 5~GeV at the PS test beam.
Similar distributions for SPS energies, for electrons from 10 to 50~GeV are shown in \Fig{fig:adcsum_SPS}.
As shown in \Fig{fig:adcsum_PS} and \Fig{fig:adcsum_SPS}, a clear separation of the ADC sum distributions for different beam energies can be seen in both PS and SPS data. 
At the lowest energies at PS and SPS, the response is somewhat asymmetric due to the pedestal noise cut.


\begin{figure}[t!]
\begin{center}
\includegraphics[width=12cm]{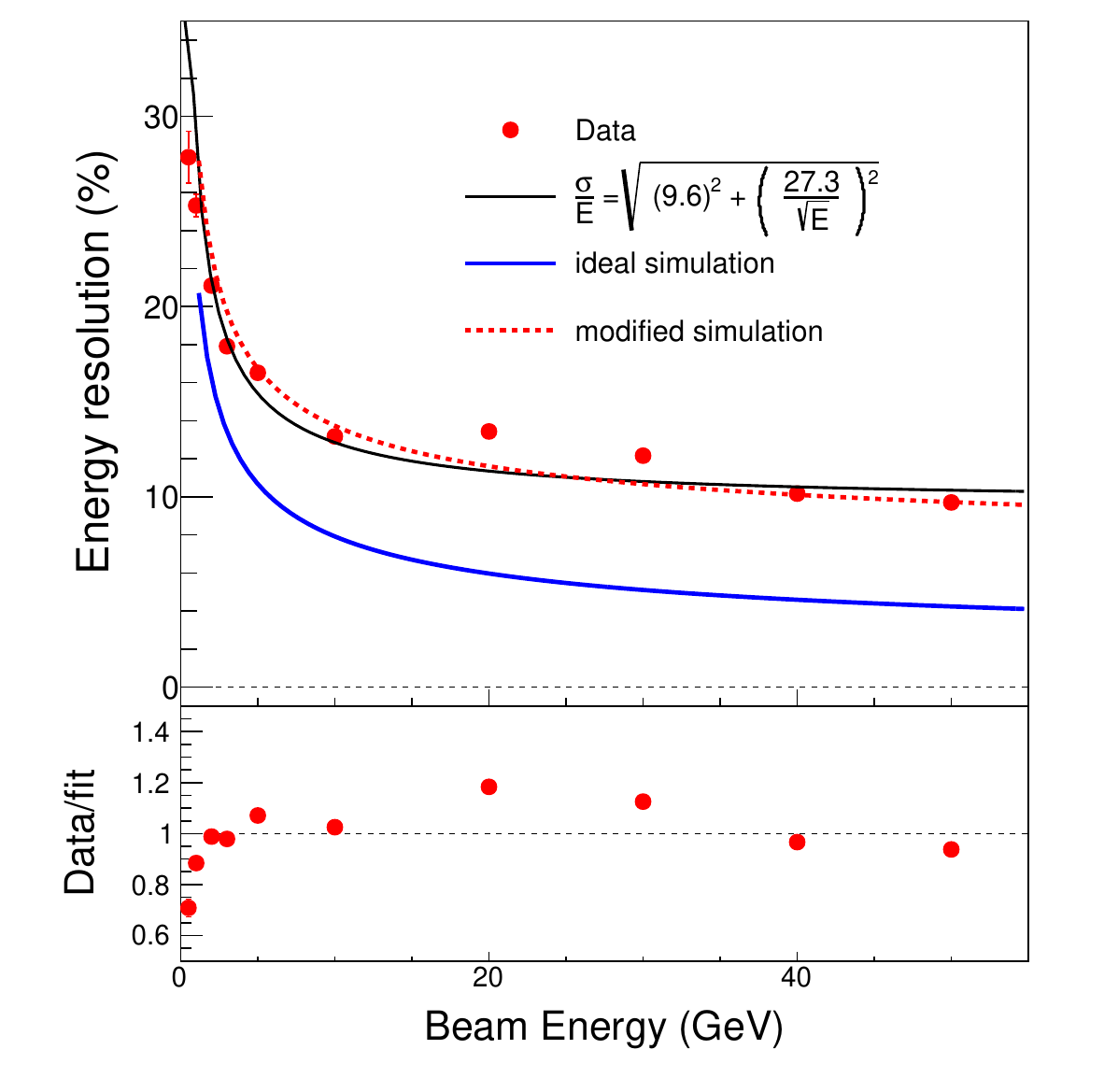}
\end{center}
\vspace{-5mm}
\caption{Energy resolution as a function of beam energy from PS to SPS data. The results of ideal simulation (solid line) and  modified simulation (dashed line) are also shown.}
\label{fig:resolution}
\end{figure}

\begin{figure}[t!]
\begin{center}
\includegraphics[width=12cm]{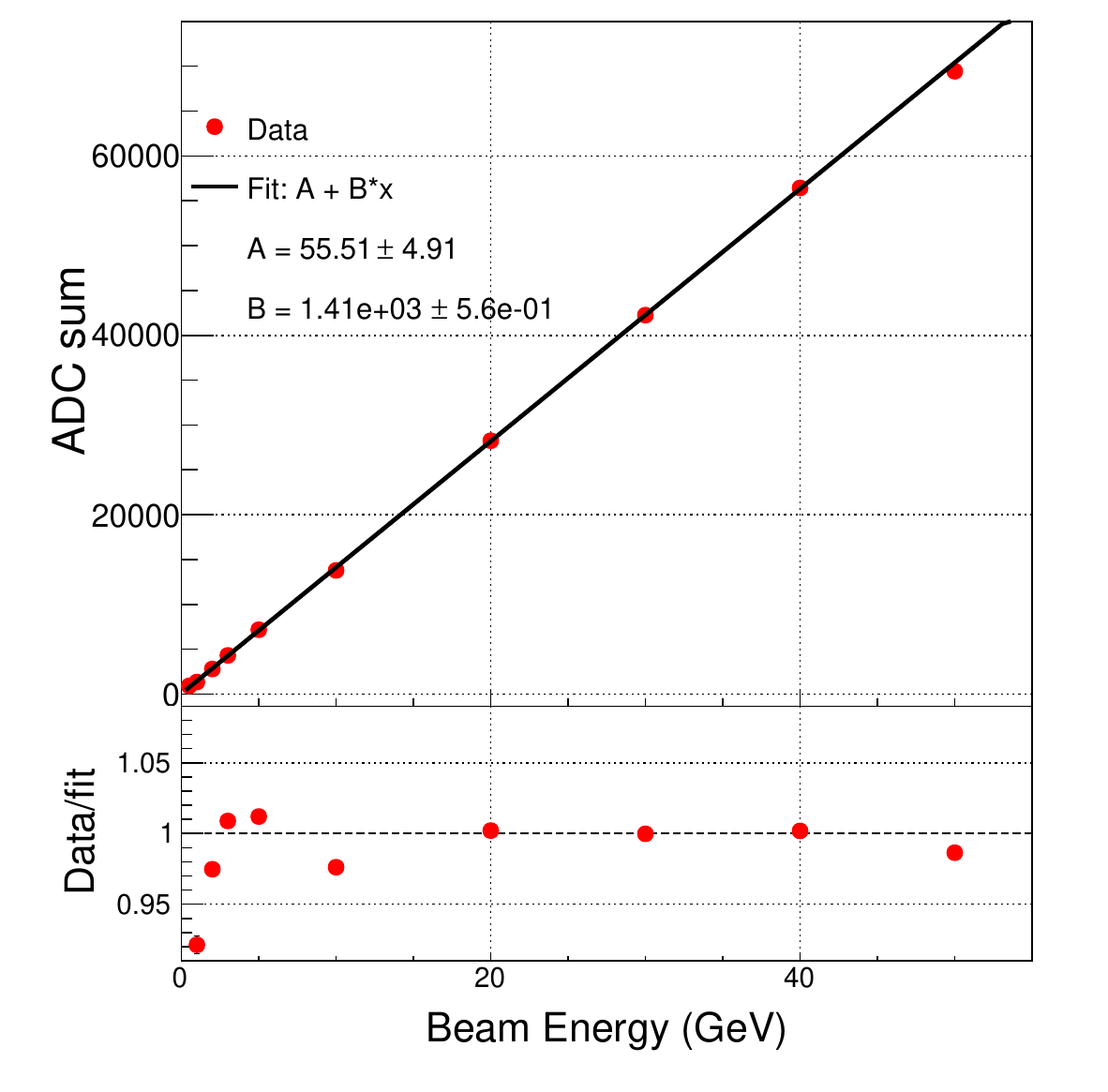}
\end{center}
\vspace{-5mm}
\caption{ADC sum as a function of incident beam energy from PS and SPS test beam.}
\label{fig:linearity}
\end{figure}

From the ADC sum distribution, we obtain the energy resolution~($\sigma/E$) and energy linearity. 
\Fig{fig:resolution} shows the energy resolution as a function of beam energy, quantified as the width of the ADC distribution $\sigma$ over the ADC mean value. 
To parameterise the energy resolution, a fit was performed to the data points with the following fit function:
\begin{equation}
    \frac{\sigma}{E} = \sqrt{B^2 + \left(\frac{A}{\sqrt{E}}\right)^2},
\end{equation}
where, $A$ is the stochastic term and $B$ is the constant term.
We obtain an energy resolution of 27.3\%/$\sqrt{E}$  for the stochastic term~(A) and about 9.6\% for the constant term~(B), as shown in \Fig{fig:resolution}.
The stochastic term is close to the expected value from the simulation.
For the constant term, a better performance  of around 1\% is expected from the ideal simulation, which however neglects some dead regions in the layers and noise in the electronics. 
In order to match the 2015 data set, in the simulation we also included a dead area corresponding to one silicon pad in the LGL2 segment located on one corner of the 3$\times$3 PAD cluster. 
Due to the trigger setup, the incoming beam particle is impinging on the detector distributed over a 1 cm$^2$ region, therefore we smeared the particle position in the simulation accordingly. 
Furthermore, realistic electrical noise, such as the pedestal width and remaining common mode noise need to be included in the simulation.
After adding the dead area as well as realistic electronic noise in the simulation, we can reproduce the data as shown in the figure. 

The energy linearity was measured as shown in \Fig{fig:linearity}, and was found to be better than 3\%, except at the lowest beam energy of 0.5~GeV. 

\section{Pixel detector performance}
\label{sec:maps}
\begin{figure}[t!]
\begin{center}
\includegraphics[width=12cm]{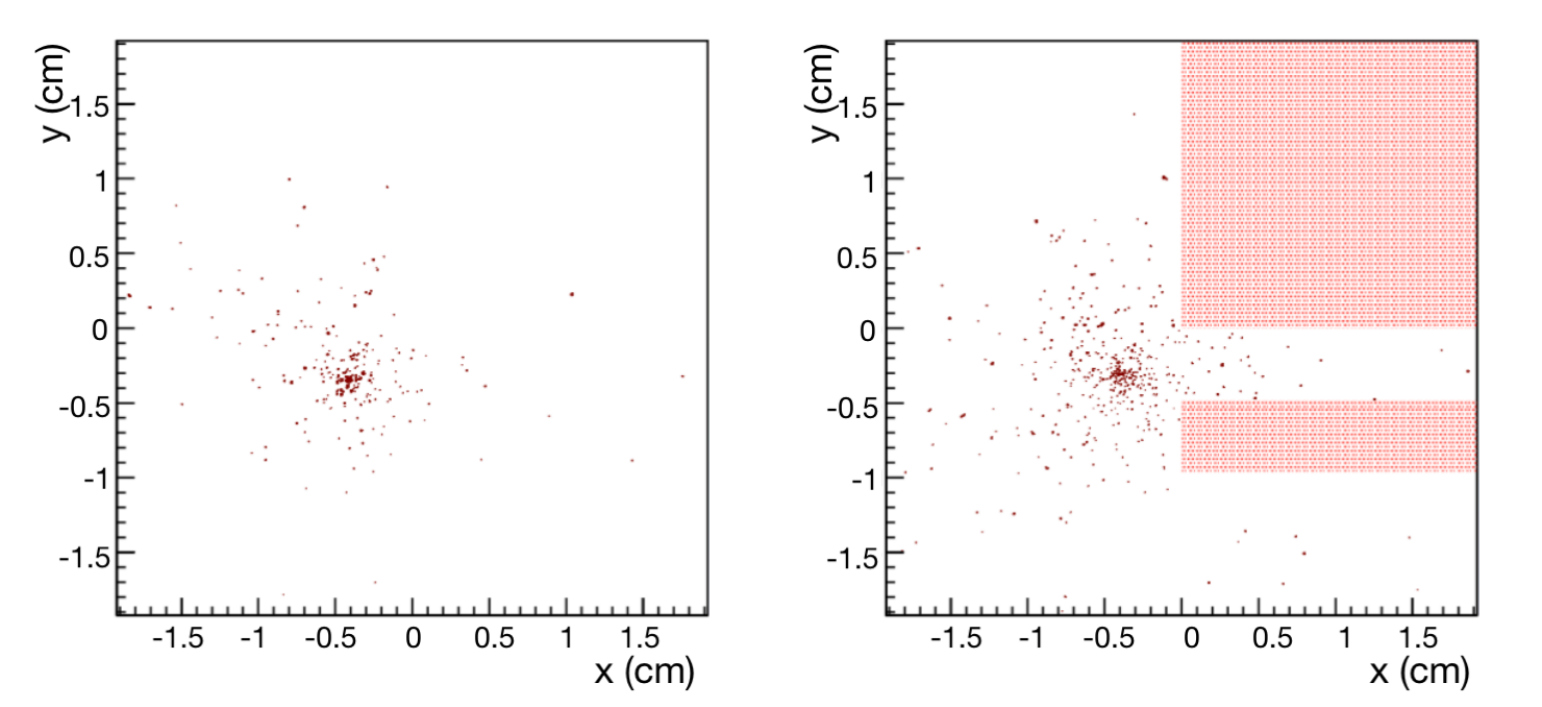}
\vspace{-5mm}
\caption{(colour online) A single event hit map for electromagnetic showers produced by 130\,GeV positron beam.
         The left panel shows the forward and the right panel the backward layer.
         The red zones in the hit maps are dead areas where we were not able to obtain signals.}
\label{fig:hitmap}
\end{center}
\end{figure}

One of the main purposes of the pixel layers is to facilitate the reconstruction of pairs of photons from neutral pion decays. To do this, they need to provide good two-shape separation and provide good estimates of the position and energy of the single showers in the pair. Measurements with a full digital pixel calorimeter prototype \cite{de_Haas_2018} have shown that the energy linearity and resolution are good and that the position resolution is excellent. With the present measurements we can verify whether the signals in two individual pixel layers provide a clear enough signal to identify showers with sufficient precision.

The data presented in this section were taken in the 2016 SPS test beam with positrons of beam energies of 50, 60, 70, and 130 GeV. For the analysis, noisy pixels are removed by masking all pixels that showed a hit frequency in measurements without beam above a predefined level.

\begin{figure}[t!]
\begin{center}
\includegraphics[width=6cm]{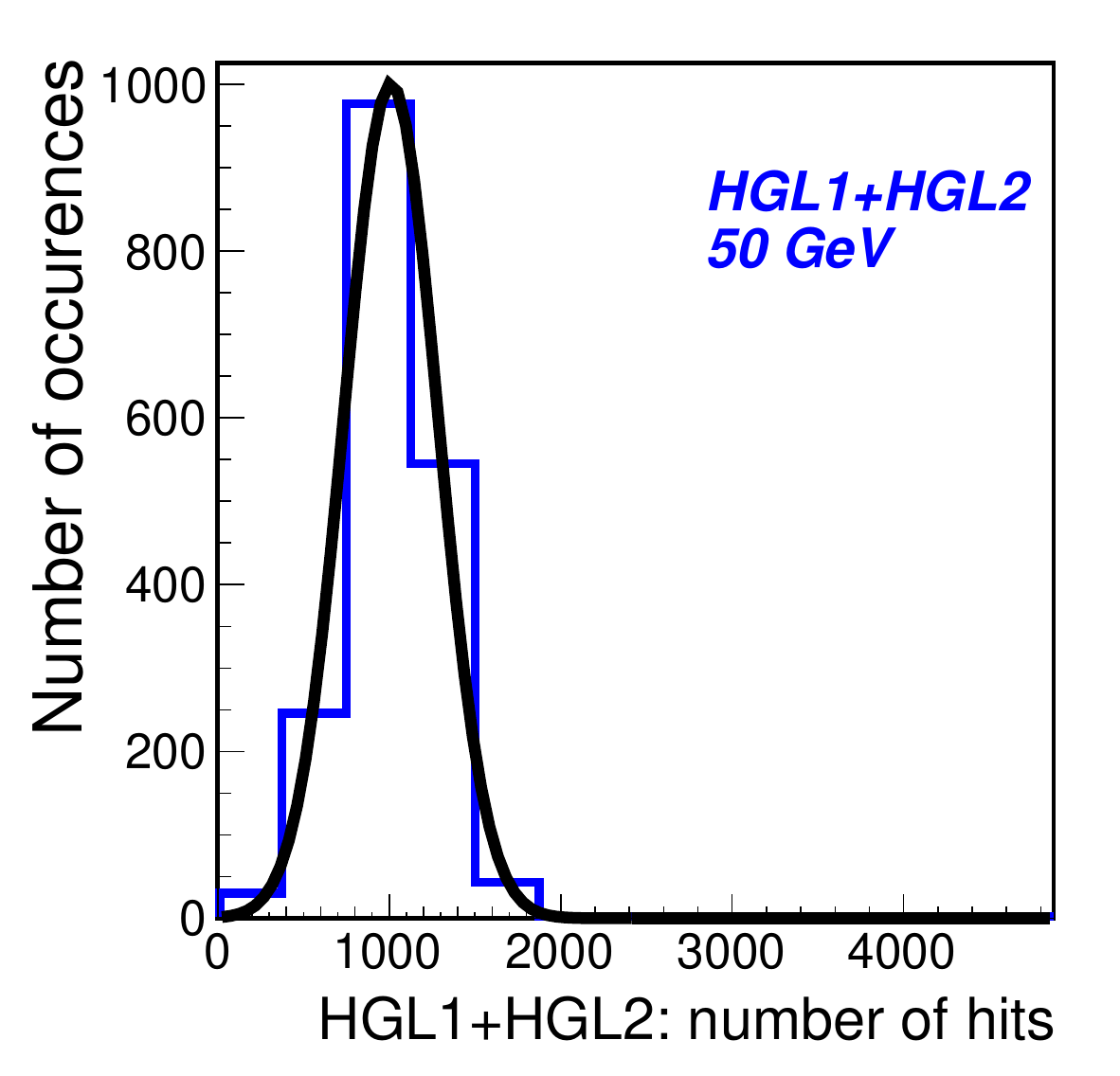}
\includegraphics[width=6cm]{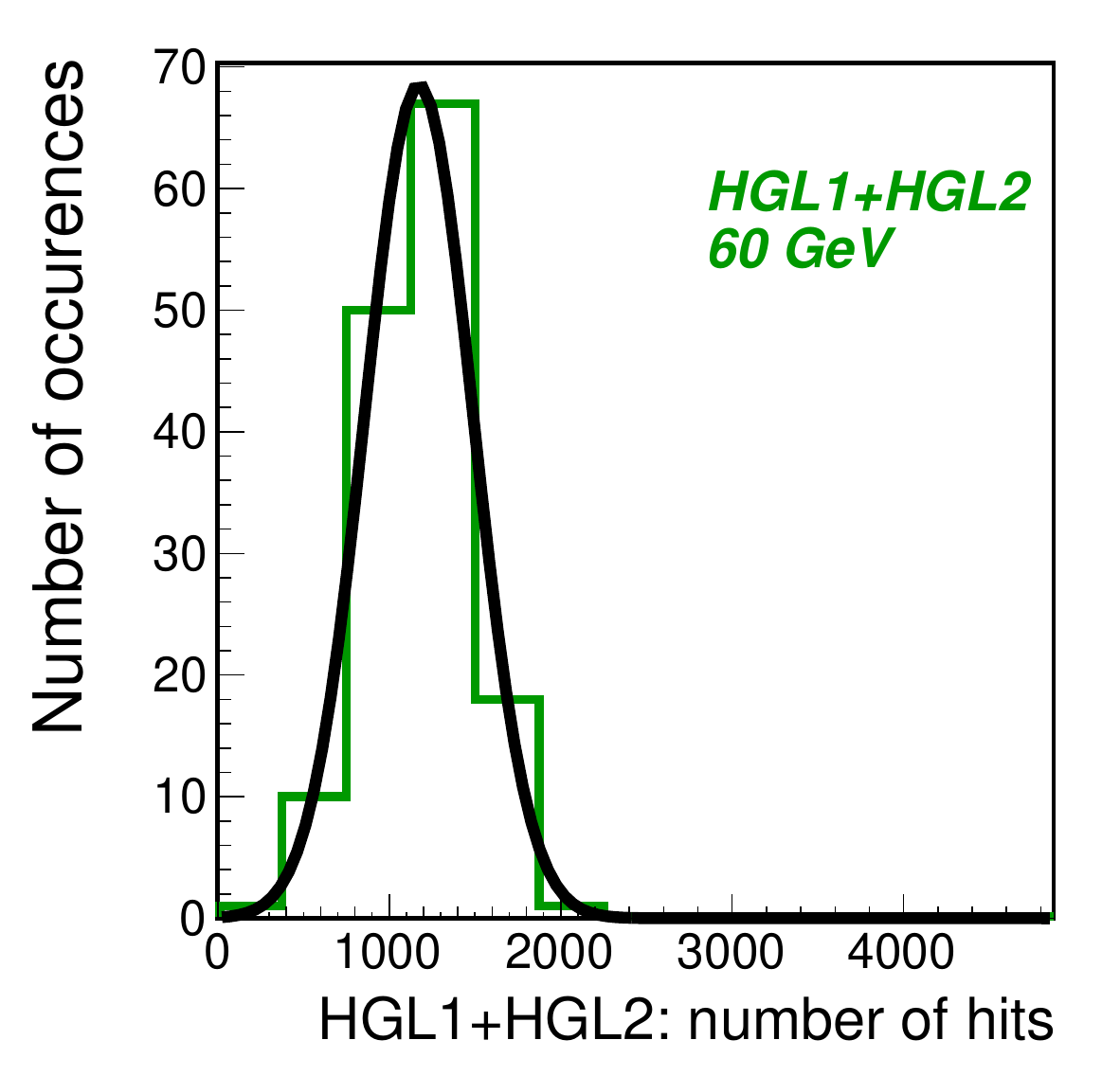}
\includegraphics[width=6cm]{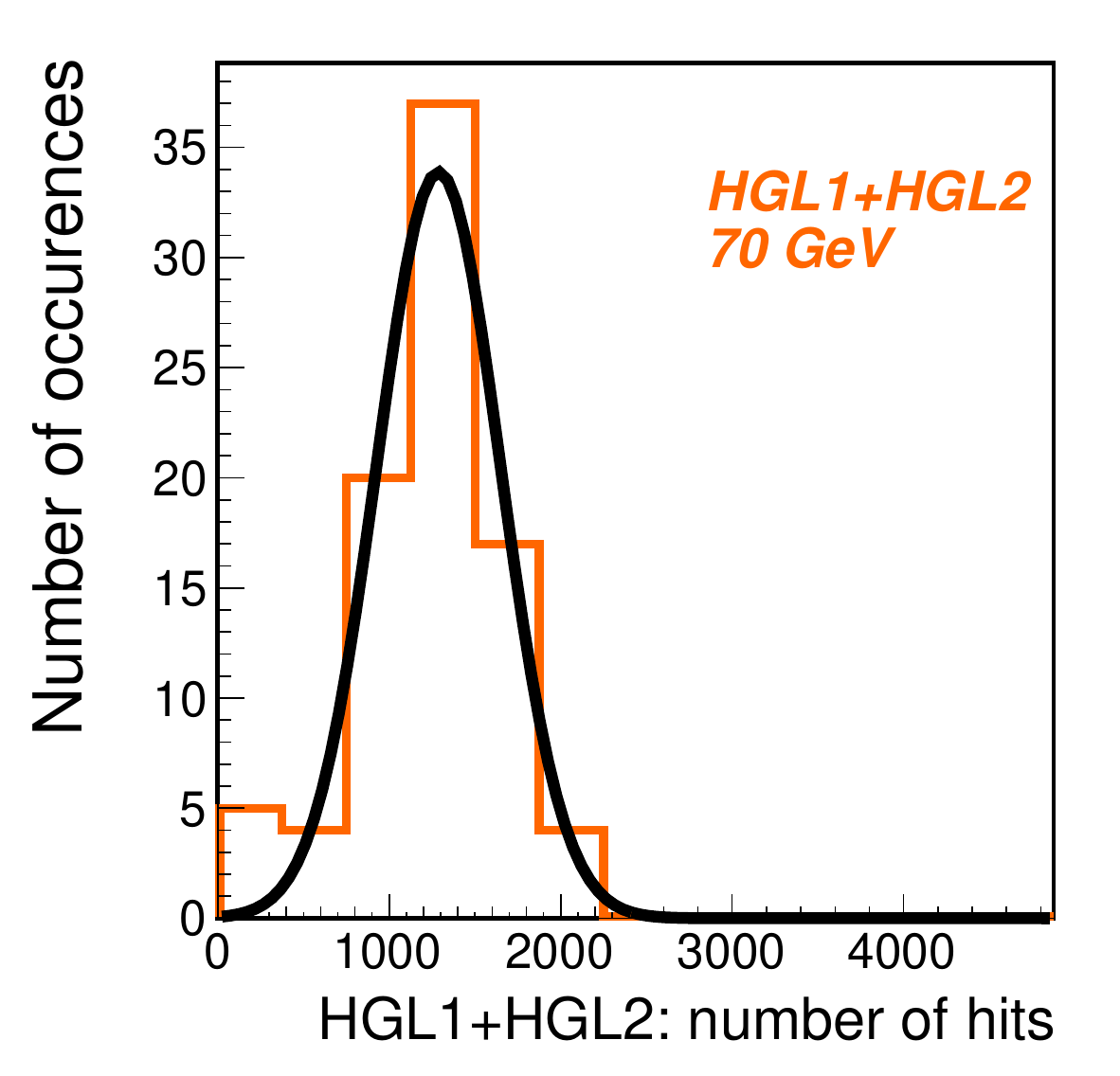}
\includegraphics[width=6cm]{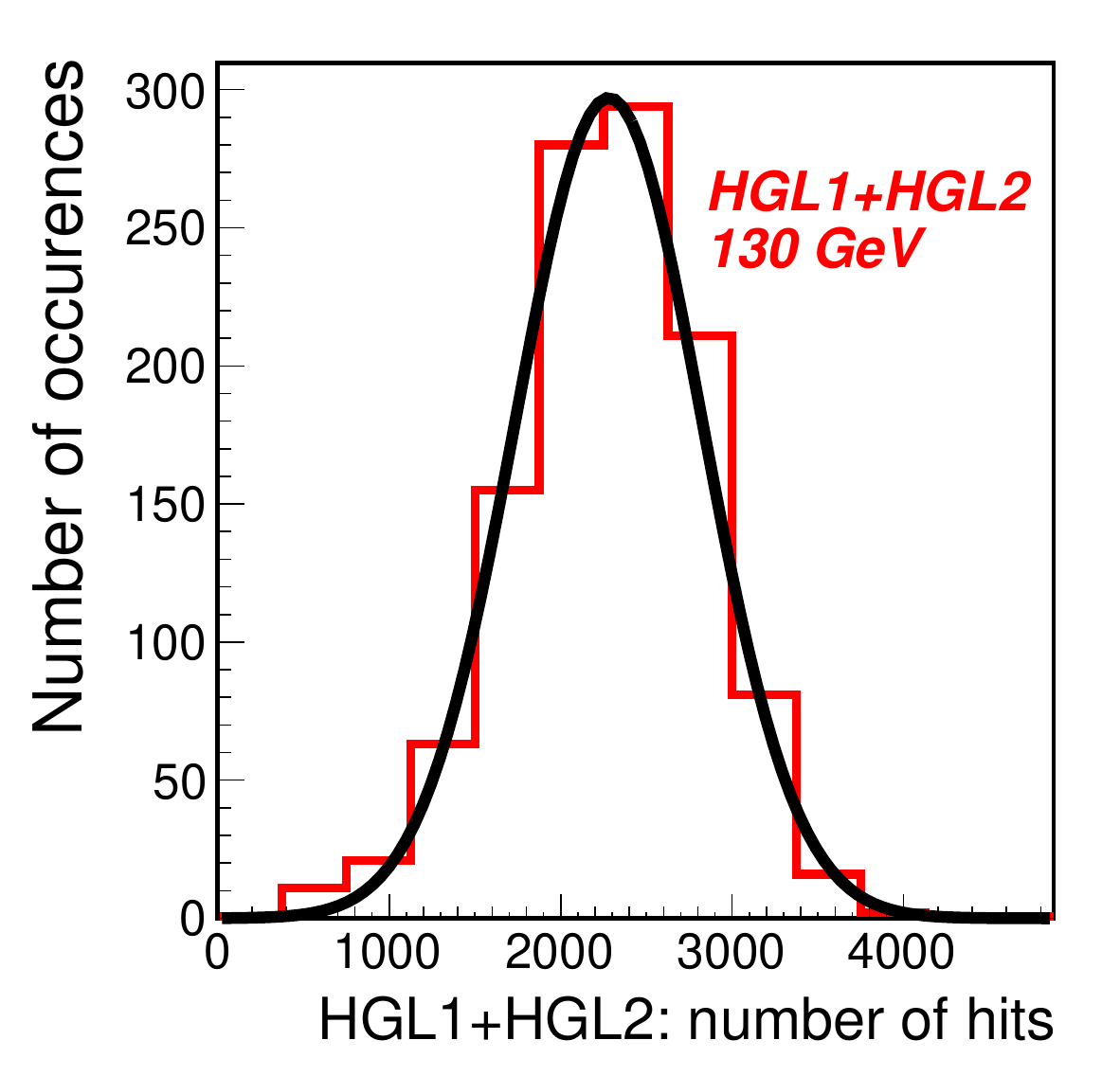}
\vspace{-5mm}
\caption{The distributions of the number of hits (sum of HGL1 and HGL2) with positron beams of 50, 60, 70, and 130\,GeV.
         Black curves show Gaussian fits.}
\label{fig:EnergyDistribution_Sum}
\end{center}
\end{figure}

\Fig{fig:hitmap} shows hit maps of a single event for an electromagnetic shower of 130 GeV positrons in the two pixel layers, which exhibit the expected shower profile in the transverse direction.
The longitudinal development of the shower from the front layer~(left) to the back layer~(right) can also be seen, where in the latter more hits are created.
The red zones in the hit map indicate dead areas of the sensors, where no hits were registered.
Those areas were excluded for the later analysis and from the calculation of the center of gravity of hits, as defined below.

\begin{figure}[t!]
\begin{center}
\includegraphics[width=12cm]{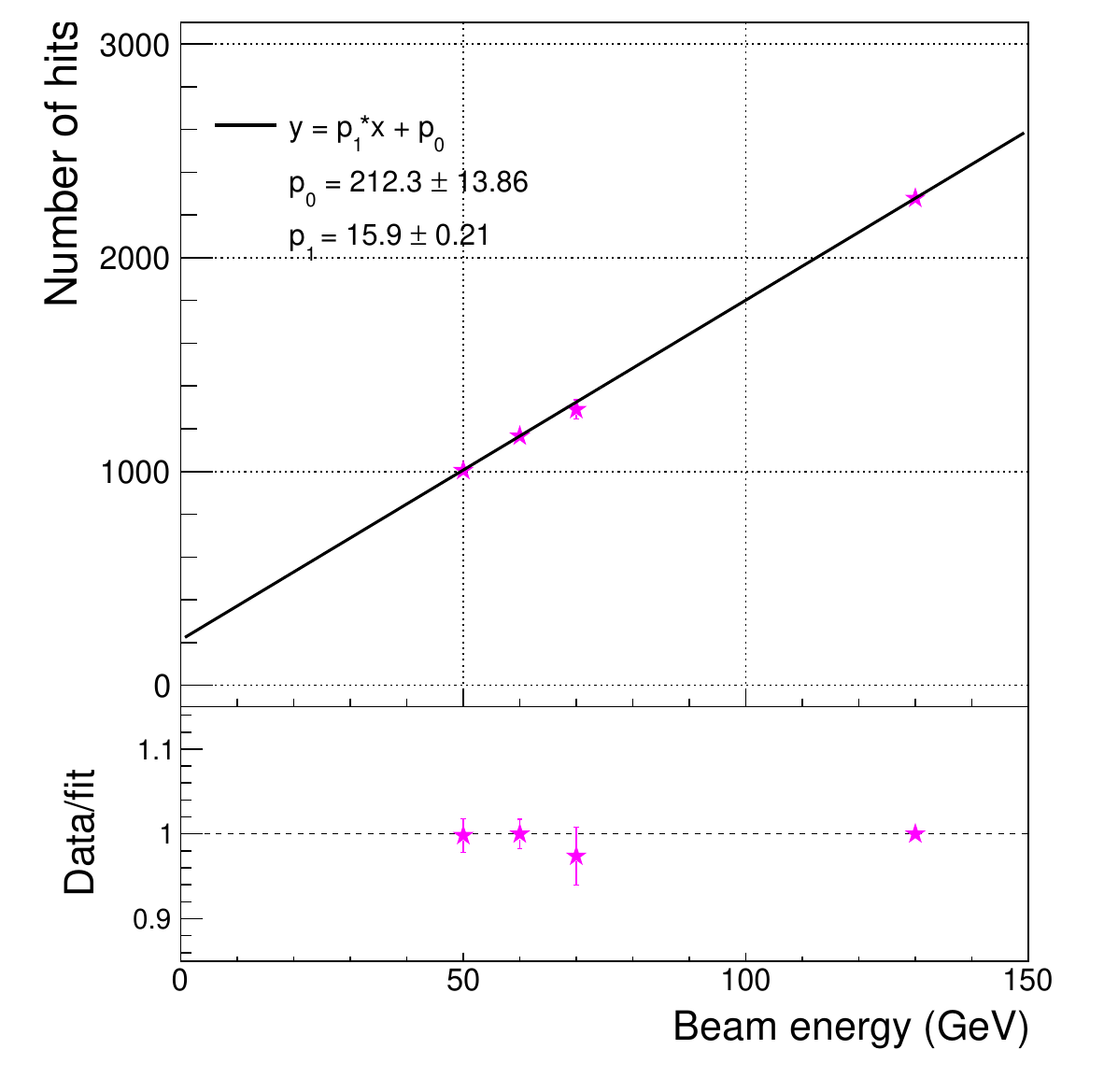}
\vspace{-5mm}
\caption{Beam energy dependence of number of hits for positrons, fitted with a linear function.}
\label{fig:Linearity}
\end{center}
\end{figure}

The shower position is determined by calculating the center of gravity in both $x$ and $y$ per event in both pixel layers separately.
We again apply a directional cut, which is now based on the pixel information, i.e.\ we select only events for which the residuals of the center of gravity of showers between two pixel layers are within $3\sigma$. 
This cut removes fake triggers and hadron punch-through events.
We also removed events which have values for the number of hits or of the RMS of the $x$ and $y$ spatial distributions which are inconsistent with the expectations for the nominal EM showers -- this mainly ensures that potential pile-up of beam particles is rejected. 
We confirmed that pile-up events were negligible according to the hit distributions.
No pad-layer information was used because the readout was completely independent for this data set.

\Fig{fig:EnergyDistribution_Sum} shows the distributions of the sum of the number of hits in HGL1 and HGL2 for positron beams of 50, 60, 70, and 130\,GeV energy.
A clear increase of the number of hits with beam energy is visible.
\Fig{fig:Linearity} shows the mean number of hits derived from those distributions as a function of beam energy. This indicates that already in the two layers used here the number of hits depends linearly on the energy and as such provides a good measure of the energy of the full shower.

\ifcomment
\begin{figure}[t!]
\begin{center}
\includegraphics[width=10cm]{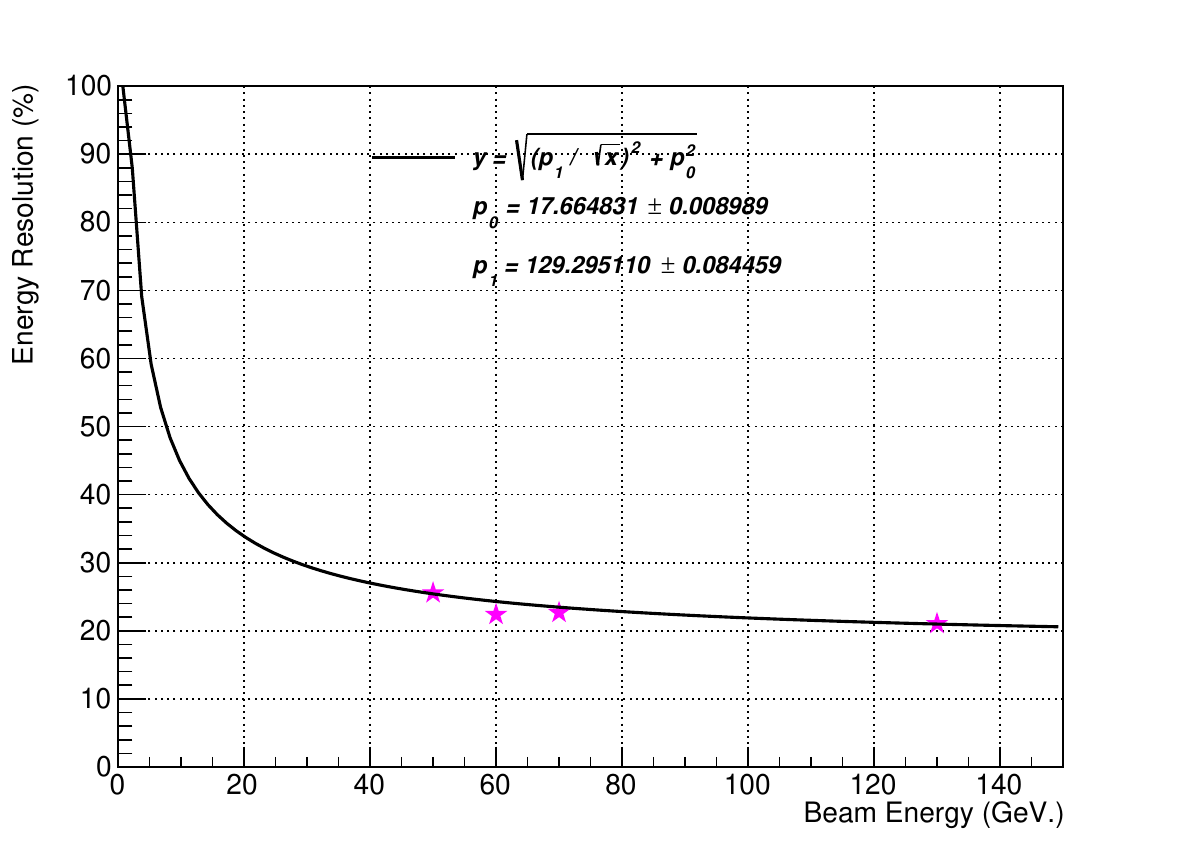}
\vspace{-5mm}
\caption{Energy resolution of the pixel layer for positron beams.
         A black line shows a fit to stochastic resolution model.}
\label{fig:EnergyResolution}
\end{center}
\end{figure}

\Fig{fig:EnergyResolution} shows the energy resolution based on the hit distributions.
The obtained energy resolution can be described by
\begin{equation}
\sigma(E) = \frac{129.295 \pm 0.084}{\sqrt{E}} \oplus 17.6648 \pm 0.0090\, .
\label{eqMAPSresolution}
\end{equation}
\rem{I'm not sure that a fit of the energy dependence with that function really makes sense here. The resolution is not good, as expected for just two layers. The values are even compatible with a constant. So the stochastic term fitted should be poorly constrained. I am surprised that the error on the parameters is so small. May be related to the fact that the error on the resolution values seems to be very small (invisible). Should be checked, but I would assume that there is something missing here. (But even if these are correct, there are too many digits specified.)
More important here is the linearity plot, which is discussed next. I think it is nice that one can at all perform an energy measurement with two layers. tp: The linearity plot is now before this, but if we leave out the resolution, this does not matter.}
The stochastic term is over a hundred, because we only have two MAPS layers in this test setup (in 4$X_0$ and 8$X_0$), so that  it is insufficient to work as a EM calorimeter.
However, these results suggest that energy of charged particles can be measured as well by counting hits.\rem{NO! This is about measuring the energy of electromagnetic showers.}
\fi

\begin{figure}[t!]
\begin{center}
\includegraphics[width=12cm]{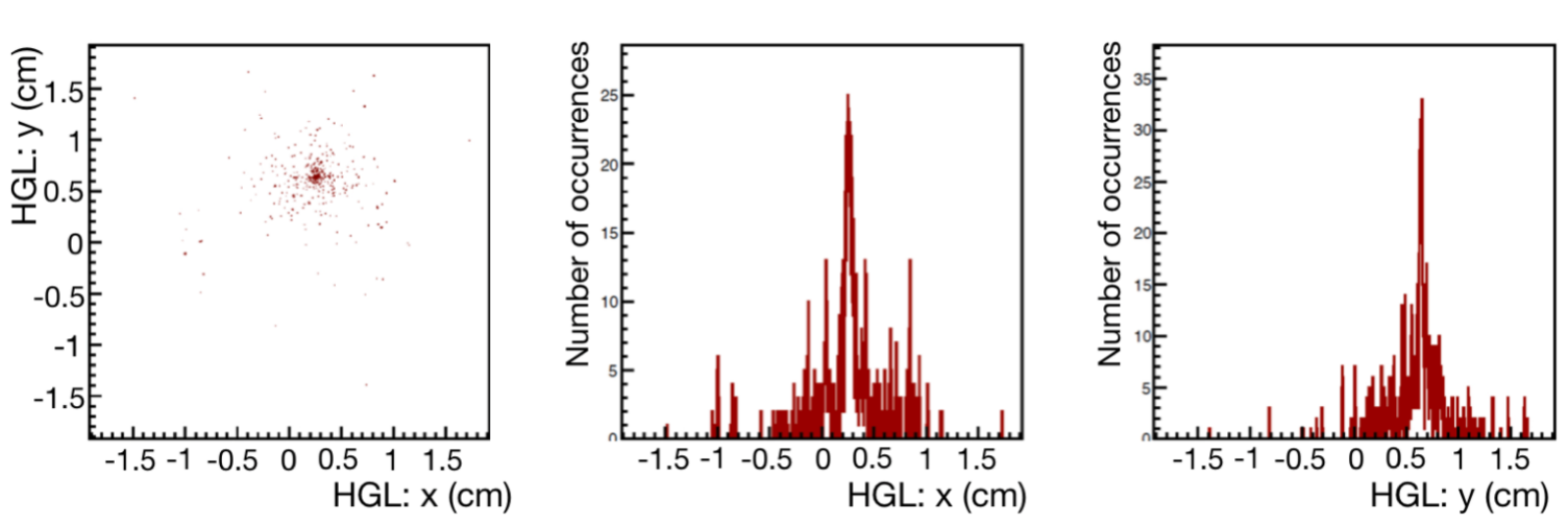}
\vspace{-5mm}
\caption{(colour online) Example of single event hit map for pixel layer~1. 
         Projection to the $x$ and $y$ axes of the hit map are also shown.}
\label{fig:gravity}
\end{center}
\end{figure}

\Fig{fig:gravity} shows the ($x$,$y$) hit positions~(left) and the corresponding $x$- and $y$-projections of the hit distributions in pixel layer 1 for a single event. 
The projections show a clear peak indicating the position of the shower center.
Most of the hits are in a very narrow region of about 1 mm around the center of the shower, which should make it possible to distinguish single photon showers from photon pairs down to separation distances of around a millimeter.

\Fig{fig:gravity_cor} shows the distribution of the center of gravity positions in $x$ from the HGL1 and HGL2 layers for different energies. 
A clear correlation between the positions in the two layers is observed, and similar behaviour is found for the $y$ direction.

\begin{figure}[t!]
\begin{center}
\includegraphics[width=12cm]{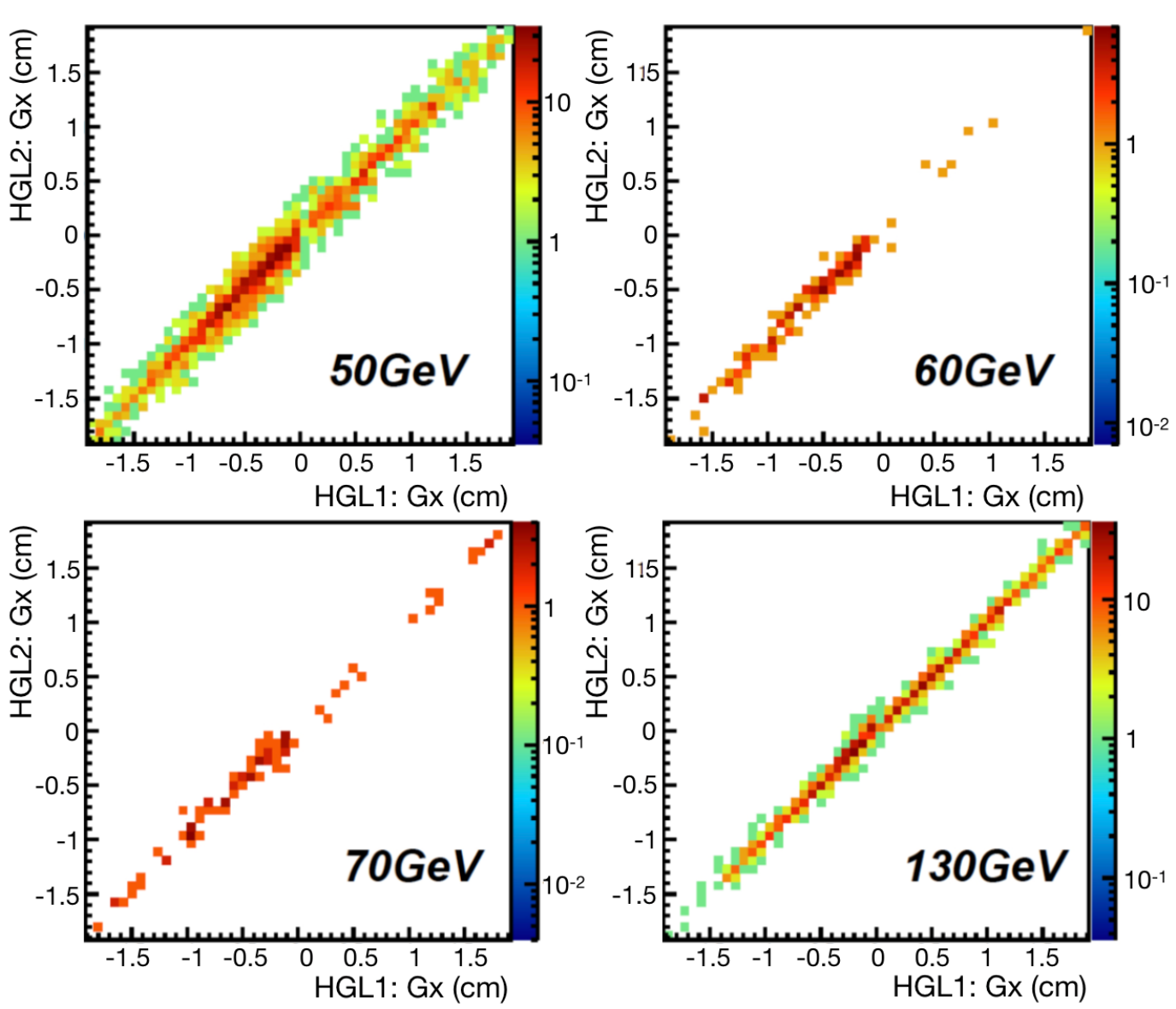}
\vspace{-5mm}
\caption{(colour online) Correlation of center of gravity for the $X$ direction between HGL1 and HGL2}
\label{fig:gravity_cor}
\end{center}
\end{figure}

\ifcomment
\begin{figure}[tbh!]
\begin{center}
\includegraphics[width=12cm]{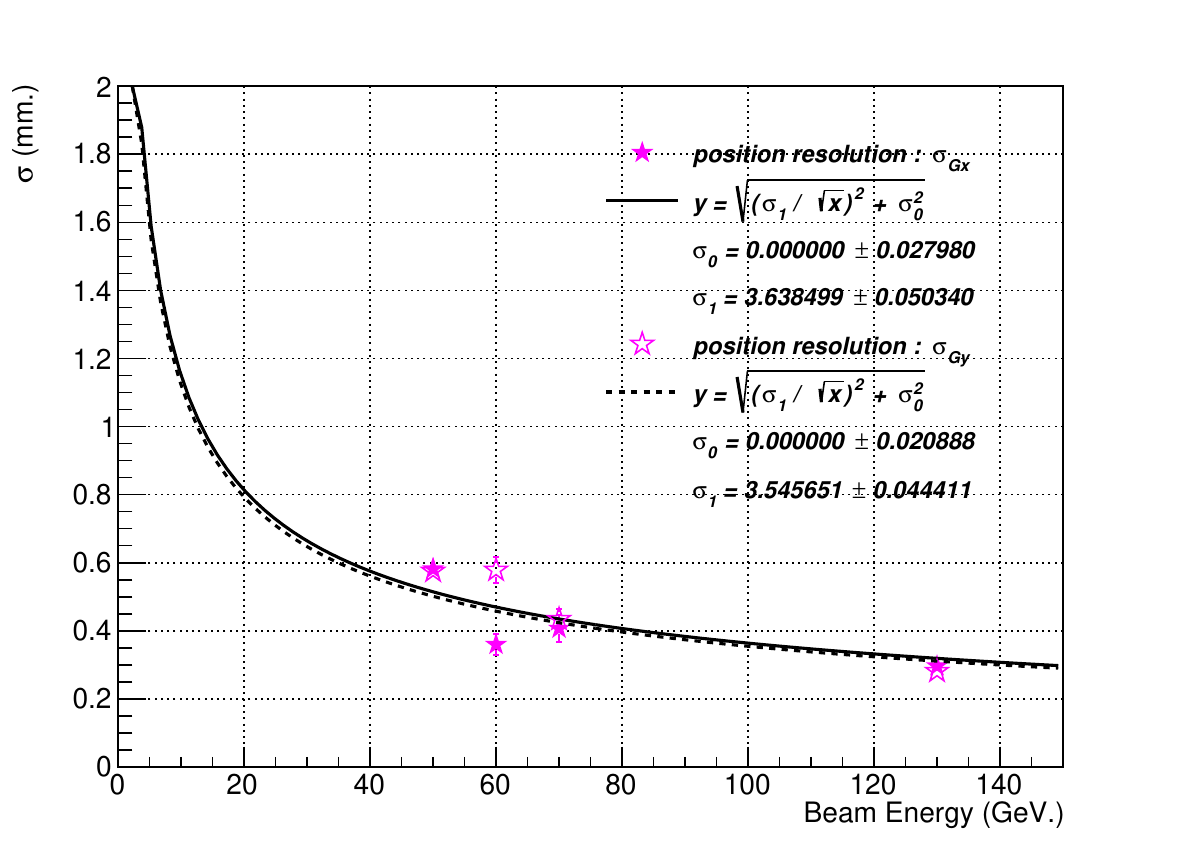}
\vspace{-5mm}
\caption{The position resolutions for different energies of positron beams in vertical plane for beam axis.
         A black line shows a fit to the resolutions in x axis , and dashed line is to the resolutions in y axis.
         Stochastic resolution model is applied as fit functions in both of axes.\rem{compare with pixel paper, I wonder whether this agrees with simulation and/or the full pixel testbeam; maybe we can dig up some information on this from a thesis from Utrecht.}}
\label{fig:PositionResolution}
\end{center}
\end{figure}
\fi

The position resolution of the determination of the shower center was estimated by using the residuals of the center of gravity between two layers for both $x$ and $y$ direction.
The width $\sigma_{\Delta G}$ is used to estimate the position resolution $\sigma_G$,
\begin{equation}
\sigma_G \simeq \frac{\sigma_{\Delta G}}{\sqrt{2}}\,,
\end{equation}
assuming that the effective position resolution is the same in both pixel layers.
\ifcomment
And measured position resolution of electromagnetic shower $\sigma_{G_{x}}$, $\sigma_{G_{y}}$ are shown in \Fig{fig:PositionResolution}.
The fit results can be described by
\begin{equation}
\sigma_{G_{x}} = \frac{(3.638 \pm 0.050 \mathrm{mm})}{\sqrt{E}} \oplus (0.000 \pm 0.028 \mathrm{mm}),
\label{eqposix}
\end{equation}
and
\begin{equation}
\sigma_{G_{y}} = \frac{(3.546 \pm 0.044 \mathrm{mm})}{\sqrt{E}} \oplus (0.000 \pm 0.021 \mathrm{mm}).
\label{eqposiy}
\end{equation}
\fi
The values for $\sigma_{G_{x}}$ and $\sigma_{G_{y}}$ were found to be consistent within uncertainties, with a value on the level of 0.35~mm, decreasing with increasing energy. As expected, this is significantly larger than the single shower position resolution of $\sigma_{x,y} < 30 \, \mu\mathrm{m}$ that was obtained with a full 24 pixel layer prototype in \cite{de_Haas_2018}. 
The resolution obtained here is sufficient to match clusters in the pixel and pad layers with good spatial resolution.
The two-shower separation power is not only determined by the position resolution, but also by the width of the shower profiles as shown in Fig. \ref{fig:gravity}; both indicate that two-shower separation at the millimeter scale is possible with the pixel layers.

\ifcomment
\Fig{fig:ShowerDistribution} shows the radial distribution of the number of hits with respect to the shower center, as determined by the center of gravity of the hits in the layer. 
The transverse shower size is found to be 10--12~mm. 
The distributions are more sharply peaked for the layer closer to the front face of the detector, i.e.\ early in the longitudinal development of the shower.
The hit density in the shower core increases with beam energy, as expected.

\begin{figure}[t!]
\begin{center}
\includegraphics[width=12cm]{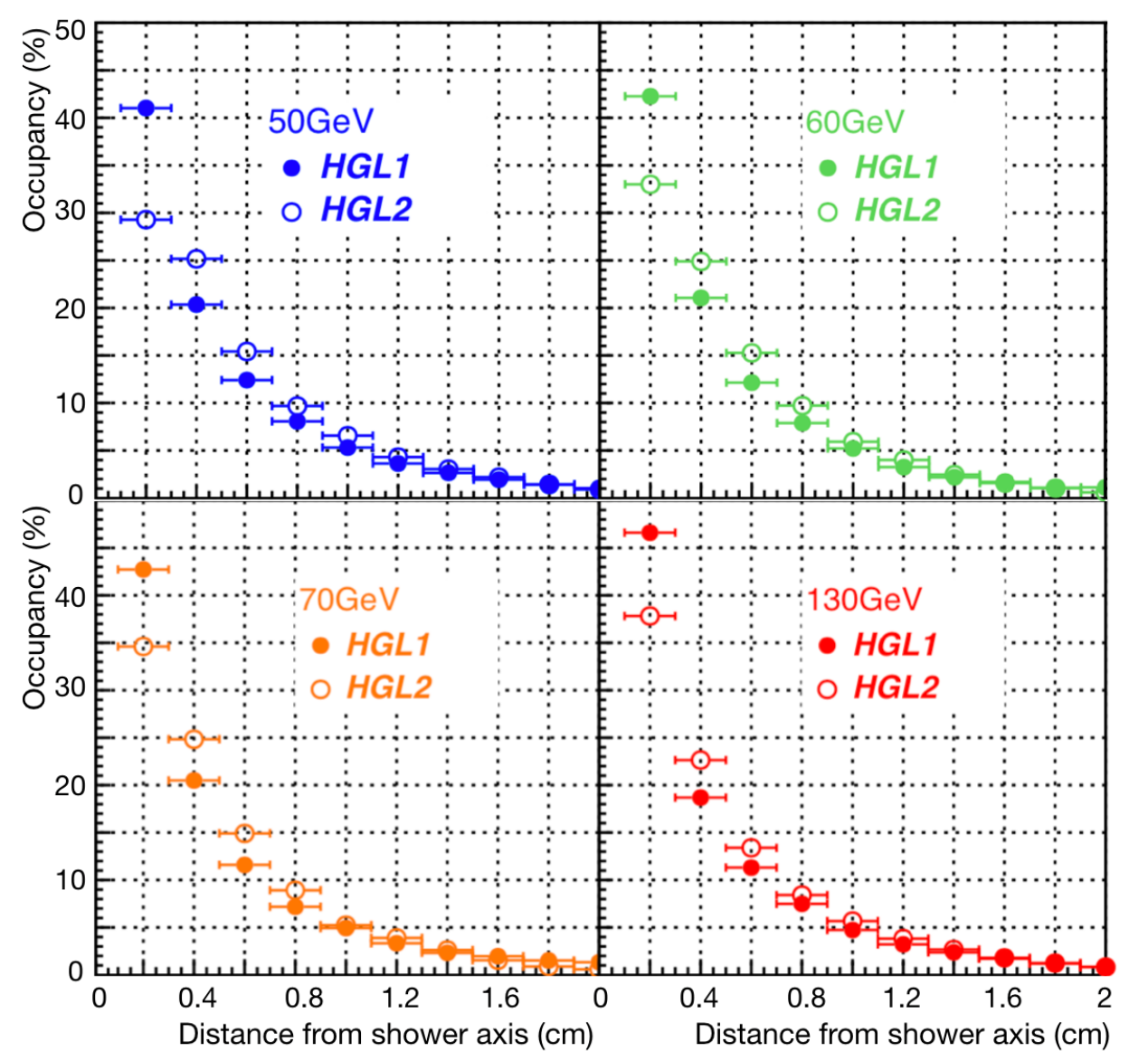}
\end{center}
\vspace{-5mm}
\caption{The transverse shower profiles for different energies of positron beams.
         Full markers show the profiles in first pixel layer, and open markers are in second pixel layer. \todo{tp: Occupancy (the y-axis labels) is not defined. I would think it should be pixel occupancy, while people might think that it is occupancy per bin in this histogram. May be one can include a proper definition and clarify that it is proportional to $dN_{hits}/dA$ - if my assumption is correct?}}
\label{fig:ShowerDistribution}
\end{figure}
\fi

\begin{figure}[t!]
\begin{center}
\includegraphics[width=11cm]{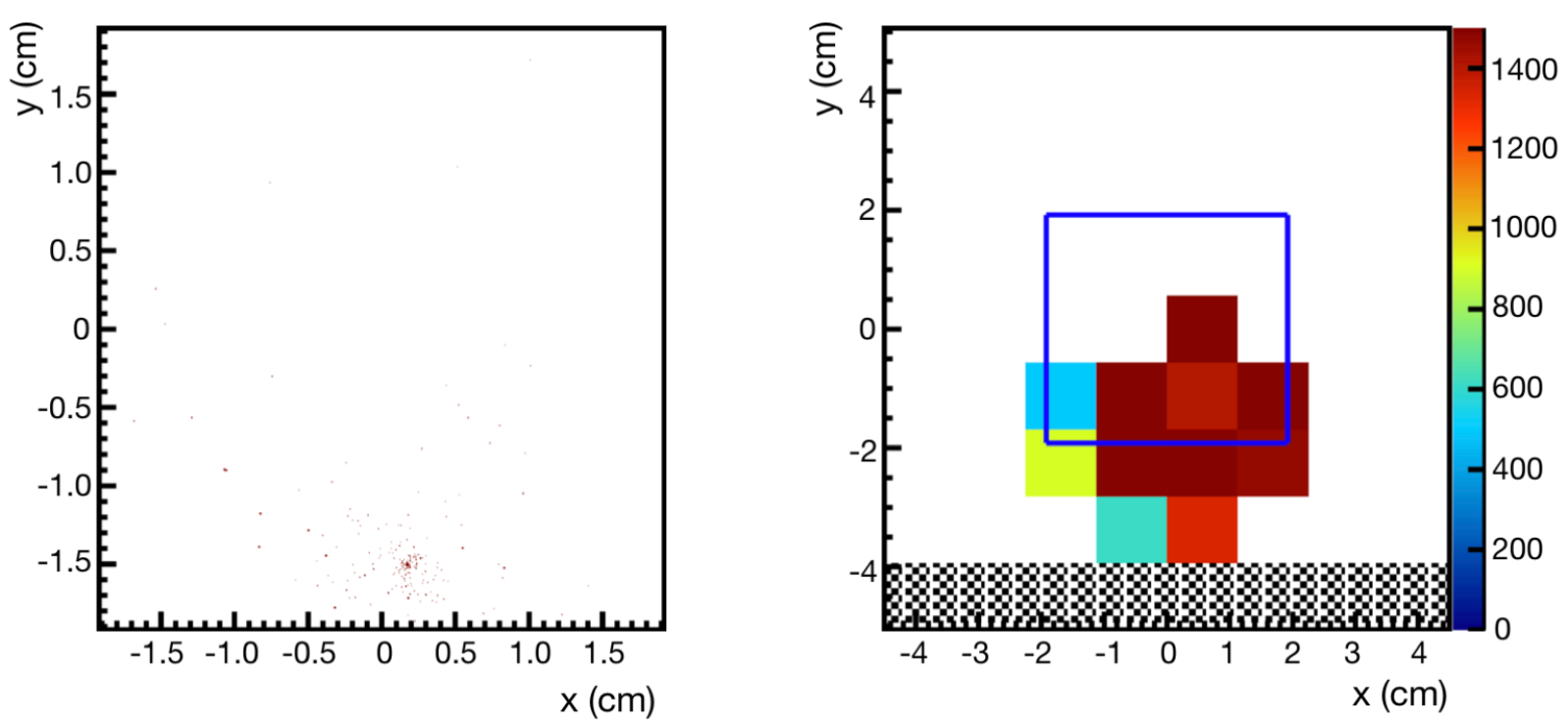}
\vspace{-3mm}
\caption{(colour online) Event display of the pixel (HGL1, left) and pad (LGL2, right) layers in an event which has common trigger information for both HGLs and LGLs.
        Black hatched area defines the region outside of the calorimeter geometry, while the blue region defines the common acceptance of the HGL1 and LGL2.}
\label{fig:matchedhitsmap}
\end{center}
\end{figure}

\begin{figure}[t!]
\begin{center}
\includegraphics[width=12cm]{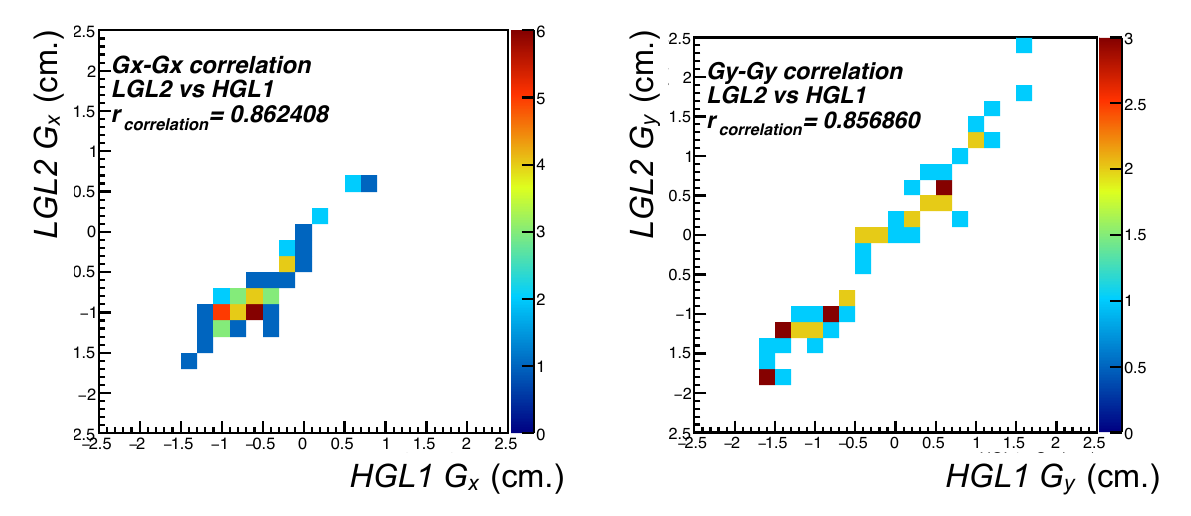}
\vspace{-5mm}
\caption{(colour online) Correlation plots for centers of showers between pixel (HGL1) and pad (LGL2) layers, with the $x$-$x$ correlation in the left and $y$-$y$ correlation in the right panel.
        The correlation factors $r_{\mathrm{correlation}}$~(see text)  are indicated (in color) on the z-axis.}
\label{fig:positioncorrelation}
\end{center}
\end{figure}

\section{Correlations between pad and pixel layers}
Although pad and pixel layers were integrated to be a common system, their detector designs are quite different from each other, and the readout systems were independent. 
Therefore to integrate hit data from pad and pixel layers, a common beam trigger given by the coincidence of two scintillation counters was stored in each data stream.
For the 2016 test beam data, we successfully obtained correlated data sets by using the common trigger info for both pad and pixel-layer data stream.
The data set was taken with 130~GeV positron beams. A hit map is shown in \Fig{fig:matchedhitsmap}. 
The left panel shows a hit map of HGL1~(pixel) and the right panel for LGL2~(pad).
The blue rectangle indicates the size of the pixel layers overlapping the pad layers.

To confirm whether the trigger matching between the data sets works correctly, the position correlation of center of gravity between the different types of layers was checked. 
\Fig{fig:positioncorrelation} shows the position of the center of gravity in one of the pad segments as a function of the position obtained from one of the two pixel layers.
The two positions are clearly correlated.
The correlation coefficient, which is given in the figure, is defined using
\begin{equation}
r_{\mathrm{Correlation}} = \frac{\frac{1}{N}
\sum_{j=1}^N (x_j-\bar{x})(y_j-\bar{y})}{\sqrt{\frac{1}{N}
\sum_{j=1}^N (x_j-\bar{x})^2}\sqrt{\frac{1}{N}
\sum_{j=1}^N (y_j-\bar{y})^2}}\,.
\end{equation}

The center of gravity in the pad layer was calculated using the ADC values as weights, which yields low accuracy due to the saturation of the readout for large signals for this particular common trigger measurement.  
However, there is a strong correlation of $r\sim 0.86$~(\Fig{fig:positioncorrelation}), which suggests that same beam events for both pad and pixel layers were successfully recorded by sharing common beam triggers.
In future analyses, we foresee making use of the position information from the pixel layers to analyse the distributions in the pad layers in more detail.

\section{Summary}
We have constructed a prototype of a silicon-tungsten sampling electromagnetic calorimeter module and the associated readout electronics, and studied performance with a test beam experiment at the CERN PS and SPS accelerator facilities in 2015/2016.
To facilitate the readout, the front-end electronics was equipped with a new custom ASIC developed for this and other application where fast, fine-granularity imaging is required. 
The energy resolution obtained is consistent with the results obtained by a realistic detector simulation.
The energy linearity is good within 3\%, from 0.5 to 50 GeV for electrons.
By sending the common beam trigger bits to both PAD and MAPS data streams, we successfully measured the 
same events in both detectors, and a clear correlation between pad and pixel layers was observed. 
\section{Acknowledgment}
We thank H.\ Muller and the RD51 collaboration at CERN for their kind support on the pad readout and frontend system using APV25 hybrid boards and SRS. 
We would also like to thank the staff members of the CERN accelerator complex for providing stable beams at PS and SPS beam tests. 
This work was supported by JSPS KAKENHI Grant Numbers JP25287047, JP17H01122, JP26220707.
Authors affiliated with ORNL are supported by the U.S. Department of Energy, Office of Science, Office of Nuclear Physics, under contract number DE-AC05-00OR22725. 
\else
\input{nimref.tex}
\fi
\section*{References}
\bibliography{mybibfile}
\end{document}